\documentclass[times,twocolumn,final,authoryear]{elsarticle}
\usepackage{jasr}
\usepackage{framed,multirow}
\usepackage{amssymb}
\usepackage{latexsym}
\usepackage{graphicx,amsmath}
\usepackage{enumitem}
\usepackage{comment}
\usepackage[switch]{lineno}

\journal{Advances in Space Research}

\begin{document}

\verso{Predrag Jovanovi\'{c} \textit{et al}}

\begin{frontmatter}

\title{The comparison of an optical and X-ray counterpart of subparsec supermassive binary black holes}

\author[1]{Predrag \snm{Jovanovi\'{c}}\corref{cor1}}
\cortext[cor1]{Corresponding author.}
\ead{pjovanovic@aob.rs}
\author[2]{Sa\v{s}a \snm{Simi\'{c}}}
\ead{ssimic@kg.ac.rs}
\author[3]{Vesna \snm{Borka Jovanovi\'{c}}}
\ead{vborka@vinca.rs}
\author[3]{Du\v{s}ko \snm{Borka}}
\ead{dusborka@vinca.rs}
\author[1,4]{Luka \v{C}. \snm{Popovi\'{c}}}
\ead{lpopovic@aob.rs}

\address[1]{Astronomical Observatory, Volgina 7, P.O. Box 74, 11060 Belgrade, Serbia}
\address[2]{Faculty of Science, University of Kragujevac, Radoja Domanovi\'{c}a 12, 34000 Kragujevac, Serbia}
\address[3]{Department of Theoretical Physics and Condensed Matter Physics (020), Vin\v{c}a Institute of Nuclear Sciences - National Institute of the Republic of Serbia, University of Belgrade, P.O. Box 522, 11001 Belgrade, Serbia}
\address[4]{Department of Astronomy, Faculty of Mathematics, University of Belgrade, Studentski Trg 16, 11000 Belgrade, Serbia}

\received{}
\finalform{}
\accepted{}
\availableonline{}
\communicated{}

\begin{abstract}
In this paper, we study and compare the optical and X-ray counterparts of subparsec supermassive black hole binaries (SMBHBs). With that aim, we simulated the profiles of optical spectral lines emitted from the broad line region (BLR) as well as X-ray spectral lines emitted from the relativistic accretion disks around both black holes and compared them with each other. The obtained results showed that SMBHBs could cause a specific, but different variability of the lines from the optical part and Fe K$\alpha$ line, leaving potentially detectable imprints in their profiles. Since these imprints depend on the orbital phase of the system, they could be used for reconstructing the Keplerian orbits of the components in the observed SMBHBs. Moreover, such signatures in the optical and X-ray line profiles of the observed SMBHBs could be used as a tool for the detection of these objects as well as for studying their properties.
\end{abstract}

\begin{keyword}
\KWD Supermassive black holes \sep variability of spectral lines \sep accretion disks\sep line: profiles
\end{keyword}

\end{frontmatter}


\section{Introduction}

The supermassive black hole binaries (SMBHB) systems are expected to be present in the centers of a number of galaxies, and they are
widely accepted that they originate in galactic mergers \citep[see][]{Begelman80,Merritt05,bogd22}.
The coalescences of their supermassive components represent the strong emitters of low-frequency (nHz - nanohertz), gravitational waves (GWs), which are currently probed by pulsar timing arrays (PTA) \citep[see e.g.,][]{ses18,chen23,liu23}, such as the International Pulsar Timing Array (IPTA) \cite{verb16}, which consists of the following three PTAs: the North American Nanohertz Observatory for Gravitational Waves (NANOGrav) \cite{mcla13}, the Parkes Pulsar Timing Array (PPTA) \cite{hobb13} and the European Pulsar Timing Array (EPTA) \cite{desv16}. Indian PTA (InPTA) \cite{josh18} and China PTA (CPTA) \cite{lee16} are also actively searching for nHz GWs, but there are also plans for even more powerful PTAs, such as the Square Kilometre Array PTA (SKAPTA) \cite{smit09} and the next generation Very Large Array (ngVLA) \cite{nanograv18}. Besides, the (active) SMBHBs will be among the most significant observational objectives of future space-based missions, such as the Laser Interferometer Space Antenna (LISA) \cite{klei16,amar23} and the TianQin \cite{luo16,luo20,mei21}. After the first observation of merging stellar mass black hole binary performed by \cite{ligo16}, searches for the SMBHB systems were significantly intensified, which resulted in more than 100 candidates until now 
\citep[see e.g.,][]{grah15,char16,ses18,agaz23}.

PTAs, ground and space-based interferometers are used for detection of GWs in different domains of frequencies: high frequency GWs (10 Hz -- 100 kHz) could be detected using ground-based detectors, low and middle frequency GWs (100 nHz -- 10 Hz) by space-based interferometers, while very low frequency GWs (less than 100 nHz) could be detected using PTAs. SMBHBs with subparsec orbital separations are among the brightest sources of low frequency GWs with frequencies of $\sim 10^{-9} - 10^{-7}$ Hz. However, individual SMBHBs emit continuous GWs, while the nanohertz GWs are expected to be in the form of stochastic background due to contributions of the entire population of SMBHBs and other potential sources. Moreover, GW background signal is expected to be detected first, while PTAs are expected to reach the sensitivities required to detect GWs from individual SMBHBs, soon after (see e.g. \cite{agaz23,nanograv23,nanograv24}). Therefore, it is of great significance to study electromagnetic counterparts of GWs emitted from individual subparsec SMBHBs, since these investigations could be used for detection of SMBHB candidates for the sources of such GWs.

Unfortunately, direct observation with current observation methods and techniques is impossible, especially in the case of close separation binaries. A potentially promising method for SMBHB detection is high-resolution radio observations \citep[see, e.g.,][]{ts13,Liu14,mo18}, but it could be applied only at the kpc scale distance \citep[see][]{fu11} between the components (black holes in binary system). However, in cases where the spectral characteristics reflect the dynamics of an orbital motion of SMBHB, spectral observations can be used for the detection of SMBHB candidates
\citep[see e.g.][]{er12,pop12,bon12,li16,wa17,ser20}. There are a number of BBH candidates reported in literature with signatures of variability in their optical curves \citep[see e.g.][]{charisi18,liu19,chen20,liao20}.

Spectroscopy in different spectral bands is one of the most powerful 
methods for SMBHB searches. For example, in a galactic merger, two SMBHs become 
gravitationally bound and start to orbit around their center of mass, which causes 
that the emission lines from SMBH components start to shift due to their radial 
velocities \citep[see e.g.][for examples in the optical band]{pop12,bon12,li16,bon19}.
Strong X-ray emission in the broad Fe K$\alpha$ line at 6.4 keV could arise from 
accretion disks around both SMBHs. Also, this strong X-ray emission could be 
affected by the Doppler shifts due to the orbital motion of the binary 
\cite{yulu01,jova14} because the radial velocities of its components could
reach $\approx1.5\times 10^4$ km/s \citep[see Table 1 in][]{pop12}. That is why it is expected
that such relativistically broadened Fe K$\alpha$ lines, as well as periodic X-ray variability, could 
be detected from very massive ($M>10^8\ M_{\odot}$) and cosmologically 
nearby ($z_{cosm} < 1$) SMBHBs \cite{sesa12}. The next generation of X-ray observatories, such as the
Advanced Telescope for High Energy Astrophysics (Athena+) \cite{nand13} and the X-ray Imaging and Spectroscopy
Mission (XRISM) \cite{xrism20}, will significantly improve the prospects for detections and detailed studies
of such SMBHB signatures in the observed Fe K$\alpha$ line profiles.

Basic dynamics and emission mechanisms of AGNs and SMBHBs are described in a number of papers \citep[see e.g.][]{osterbrock93,deRosa19}.
The spectra of active galactic nuclei (AGNs) are often characterized by a wealth of emission lines with different profiles and intensity ratios \cite[see][]{colin87,colin90,ferland1989,korista04,bruhweiler08,lamu17}. The AGN spectral line shapes in the UV and optical spectra can indicate the presence of SMBHBs 
\citep[see, e.g.,][]{pop12,nguyen16,popo21,sim2022,zh2023} and the open question: is it possible to observe similar effects in the line profiles in the X-ray and UV/optical bands?

To answer this question, we modeled the spectral line profiles in both wavelength regions: X-ray, considering the emission of the Fe K$\alpha$ line, and UV/optical, considering the emission of the H$\beta$ broad line. Our goal was to investigate the connection between X-ray and optical emission from the SMBHBs, which could be candidates for low-frequency GW sources. In general, there is a lack of information about such connections between these two emission bands, which was one of the motivations for our present study.

For the optical line emission, we use a non-hydrodynamical SMBHB model based on Keplerian kinematics, taking into account the empirical relations between dynamical parameters and emission rate \citep[][]{popo21,sim2022}. For simulation of the Fe K$\alpha$ line emitted from the SMBHB system, we used the model described in \cite{jova14} which assumes the emission of two accretion disks that each emits a broad Fe K$\alpha$ line.

Additionally, proposed SMBHB structure could be improved by including physical mechanisms which drive accretion disk winds and jets. A number of studies \citep[see][]{blan77,blan82,Begelman83,krolik01,proga04,fukumura10,nguyen19} suggests existence of such features in quasars total SED, which are superposed to the emission from the rest of the AGN. Jets are usually formed at parsec distance from accretion disk and contribute emission through synchrotron processes in radio domain, which is out of scope of this paper. On the other hand, accretion disk winds generated by any possible mechanism could in principle generate significant emission contribution in optical and X-ray domain and consequently induce variability in the observed spectrum. However, those changes are aperiodic in nature and do not reflect orbital dynamics of binary system, which is studied in this paper. Additionally, molecular and ionized winds have very long time scales of the order of few hundreds to millions of years  \citep[see][]{faucher_giguere12, fabian12, richings18}, which are much longer than the $P$, and therefore does not significantly contribute to the studied variability of the AGN in orbital time domain. Therefore, in this simple approach, we proceed with contributions of accretion disks and BLRs as a main source of emission in optical and X-ray domain, neglecting
the influence of jets and winds.

Used method is based on spectroscopic analysis, already proposed by other authors \citep[]{boro09,Begelman80,gaskel83}. System configuration corresponds to the physical picture where one or both SMBHs are active, with their BLRs possibly truncated due to BHs proximity. Larger truncation is discussed in papers of \citep{roeding14, runnoe15}, with final stage of BLR merging presented in paper of \citep{krolik19}.  In our approach we found very adoptable solution by implementing Roche lobes for BLR dimensions (see text in \S \ref{sec:opt_var}), for compact systems with separation comparable or smaller than the BLR size.

In spectroscopic search for SMBHB in SDSS catalogue, \citep{ju13} rule out existence of massive, order of $10^9\mathrm{M_\odot}$ binaries at separation $\le 0.2\mathrm{pc}$ and found that 16\% of quasars in their sample host binary system with separation $<0.1\mathrm{pc}$, but with very uncertain BH mass and BLR size. 
Our research is partially limited to the SMBHB with high eccentricity and/or inclination angle, and medium subparsec separation of $0.01-0.1\mathrm{pc}$.

This paper is organized in the following way: A model of the Keplerian barycentric orbits of SMBHBs is briefly described in Section 2, the optical and X-ray counterparts of SMBHB are discussed in Section 3. The main results of our study of the optical and X-ray variability of line profiles is presented in Section 4, and in Section 5, we summarize the obtained results.

\section{Dynamics of SMBHBs in our model}

In order to model the Keplerian barycentric orbits of a binary system of SMBHs, we used the same
procedure which is commonly applied in the case of the close binary stars \citep[see e.g.,][for more details]{hild01}. In this approach, we can assume either both masses of the primary and secondary
components $m_1$ and $m_2$, or just the primary mass $m_1$ and the mass ratio $q$ between the
secondary and primary: $q=\dfrac{m_2}{m_1}$. Besides, it is also necessary to adopt some separation
between the components, i.e. the semimajor axis $a$ of their relative orbit. Then, according to
the third Kepler's law, the orbital period $P$ of the binary is given by:
\begin{equation}
\label{eqn:period}
P^2 = \dfrac{4{\pi ^2}{a^3}}{G\left(1 + q\right)m_1}.
\end{equation}
The semimajor axes $a_1$ and $a_2$ of the true barycentric orbits (i.e. those in the orbital plane)
of the primary and secondary SMBHs are then given by:
\begin{equation}
\label{eqn:semiaxes}
a_1=\dfrac{q\,a}{1+q},\qquad a_2=\dfrac{a}{1+q},
\end{equation}
while the orbits themselves are defined in polar coordinates $(r,\;\theta)$ by the following two
ellipses whose orientations differ by 180$^\circ$:
\begin{equation}
\label{eqn:torb}
r_{1,2}\left(\theta\right)=\dfrac{a_{1,2}\left(1-e^2\right)}{1+e\cos{\theta}},
\end{equation}
where $\theta$ is true anomaly and $e$ is orbital eccentricity. Using the remaining three Keplerian
orbital elements (orbital inclination $i$, longitude of the ascending node $\Omega$ and longitude
(or argument) of pericenter $\omega$), these true orbits can be projected to the observer's sky plane
in order to obtain the corresponding apparent orbits of the primary and secondary SMBHs \citep[see e.g.][for more details]{jova20}.

Radial velocities of the components are given by the expression
\citep{hild01}:
\begin{equation}
\label{eqn:vrad}
V_{1,2}^{rad}\left(\theta\right) = {K_{1,2}}\left[ {\cos \left( {\theta  +
\omega } \right) + e \cdot \cos \omega } \right] + \gamma,
\end{equation}
where $K_{1,2}$ represents the semiaplitudes of the velocity curves:
\begin{equation}
\label{eqn:samp}
K_{1,2} = \dfrac{{2\pi {a_{1,2}}\sin i}}{{P\sqrt {1 - {e^2}} }},
\end{equation}
and $\gamma$ is systemic velocity (in our simulation it is assumed to be 0 km/s).

\begin{figure}[h!]
\centering
\includegraphics[width=0.5\textwidth]{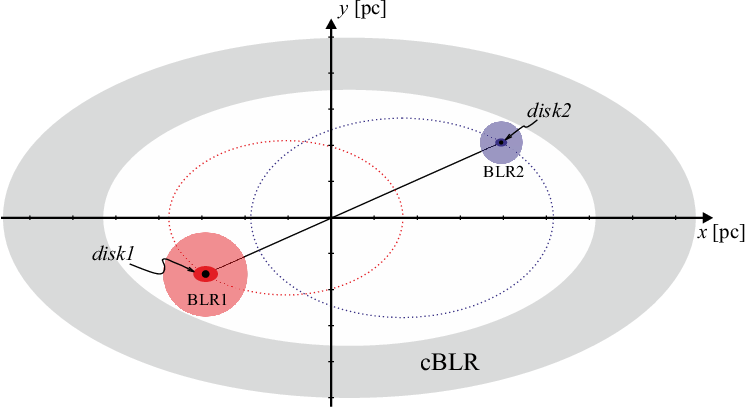}
\caption{Geometry of utilised model, gives general assumed model structure with BL regions, accretion disks and orbital paths.}
\label{fig:model_geomerty}
\end{figure}

In Fig. \ref{fig:model_geomerty} we give graphical presentation of model geometry. It distinguishes assumed general structure of the SMBHB system containing orbital paths and considered emission regions which include accretion disks and BLRs for both components, as well as surrounding circum-binary (cBLR) region. Image is shown with non zero inclination angle.

\begin{figure*}[ht!]
\centering
\includegraphics[width=0.71\textwidth]{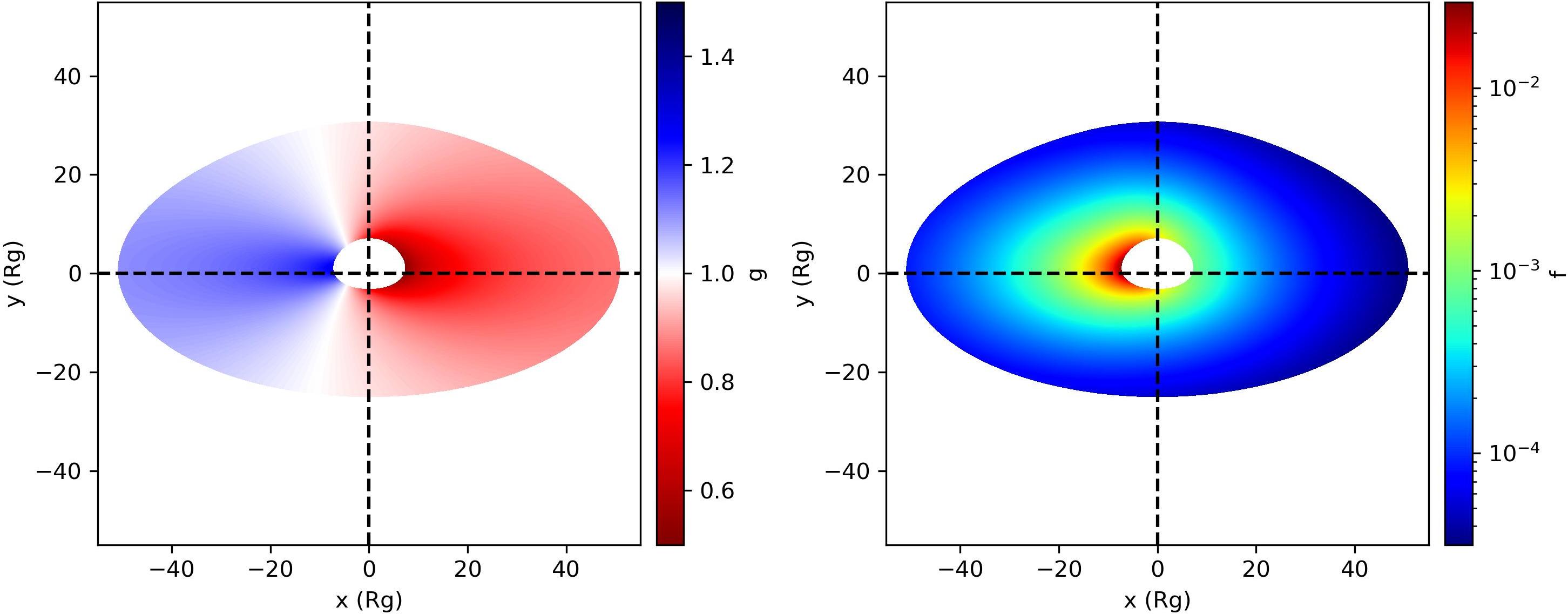}
\hfill
\includegraphics[width=0.28\textwidth]{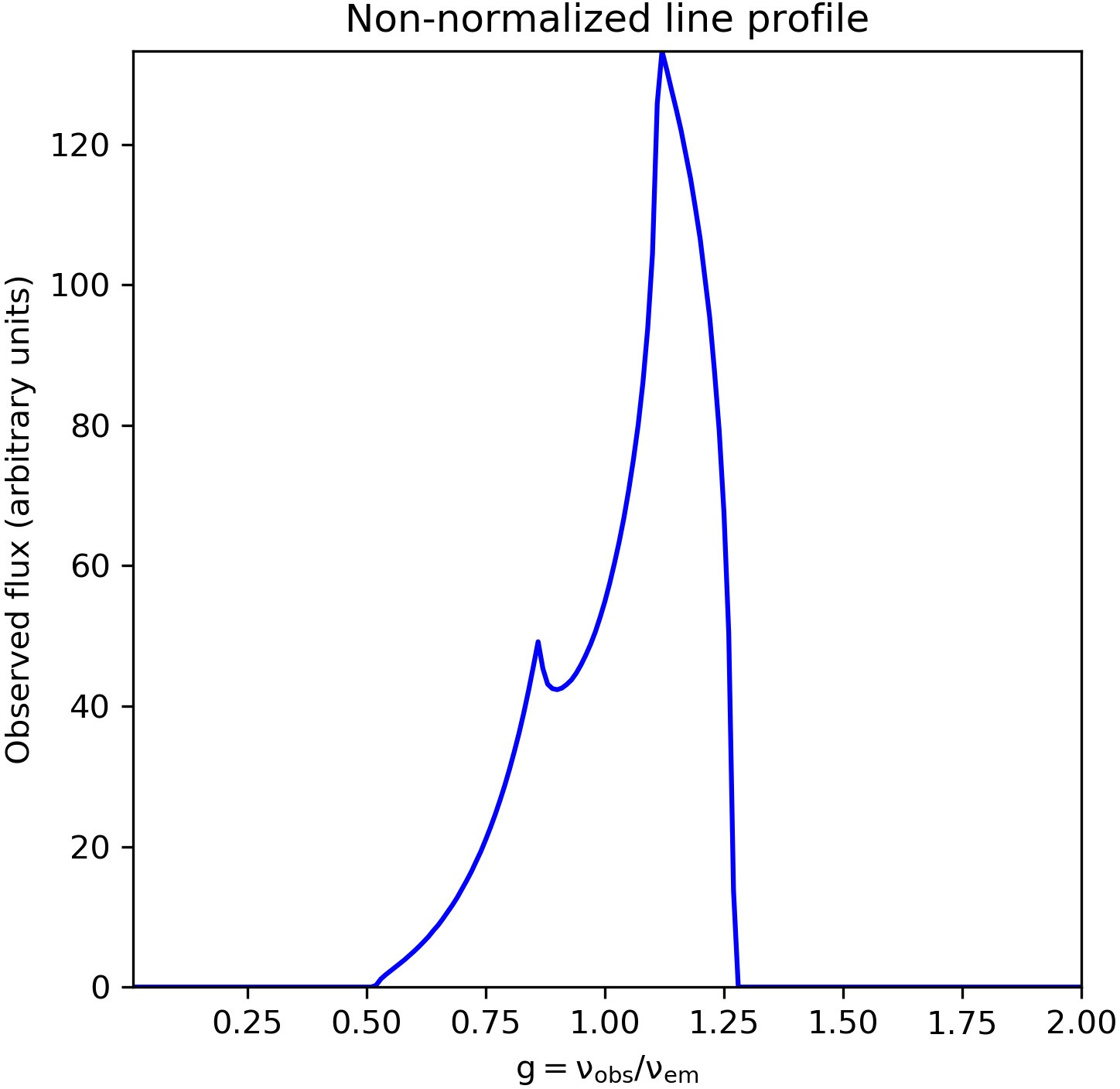}\\
\includegraphics[width=0.71\textwidth]{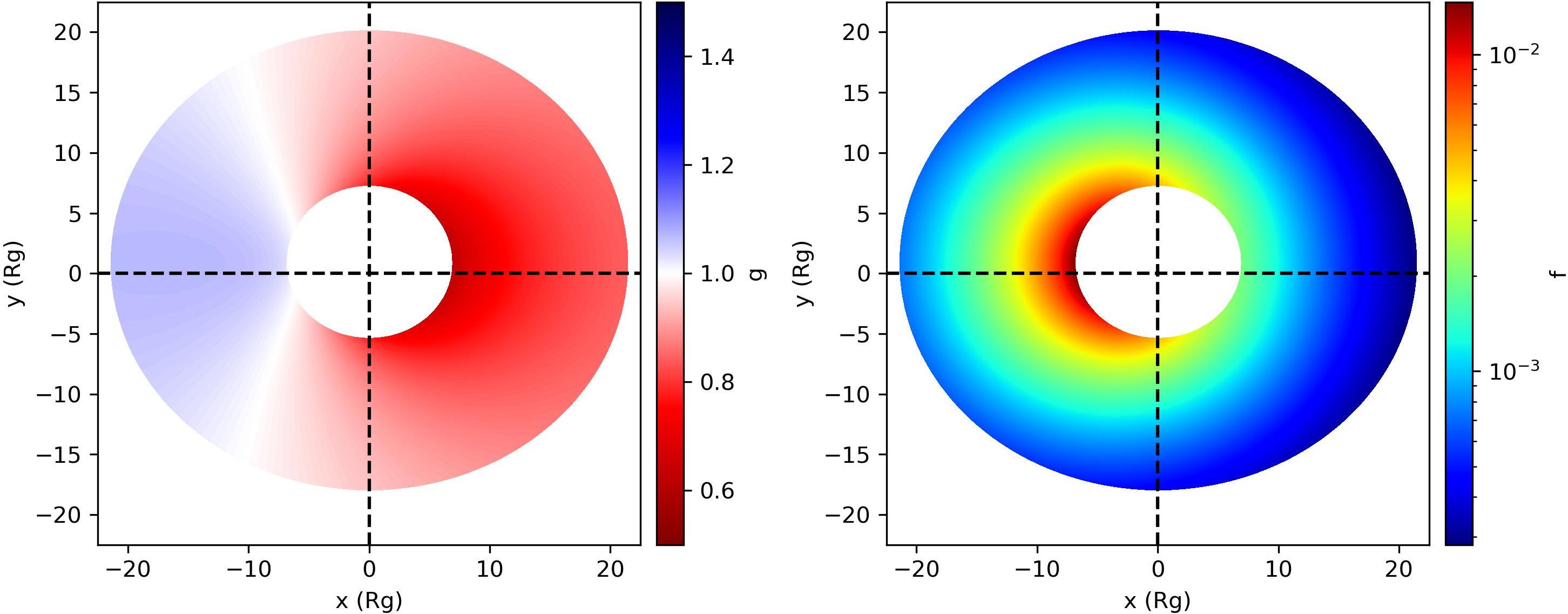}
\hfill
\includegraphics[width=0.28\textwidth]{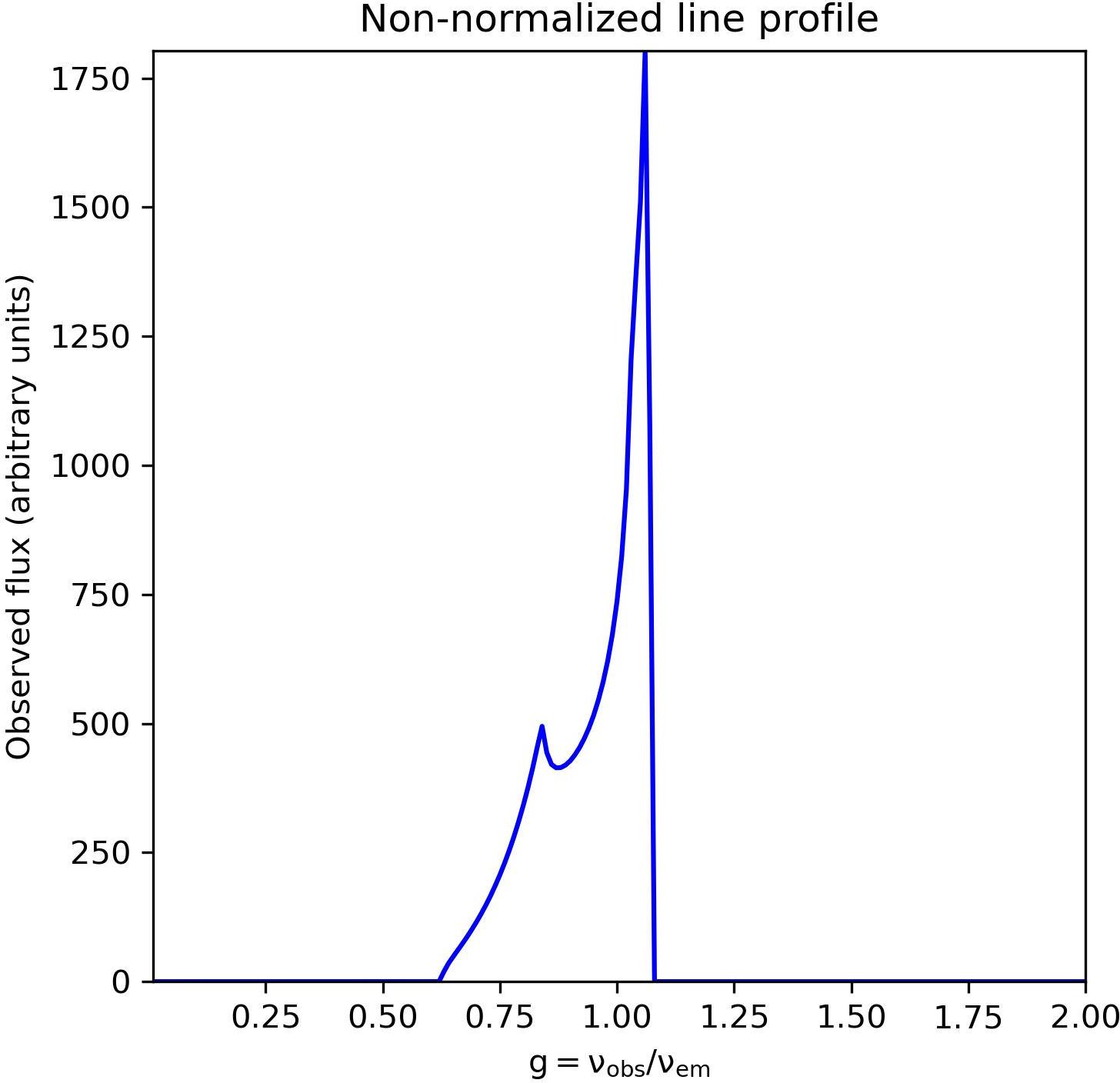}\\
\includegraphics[width=0.71\textwidth]{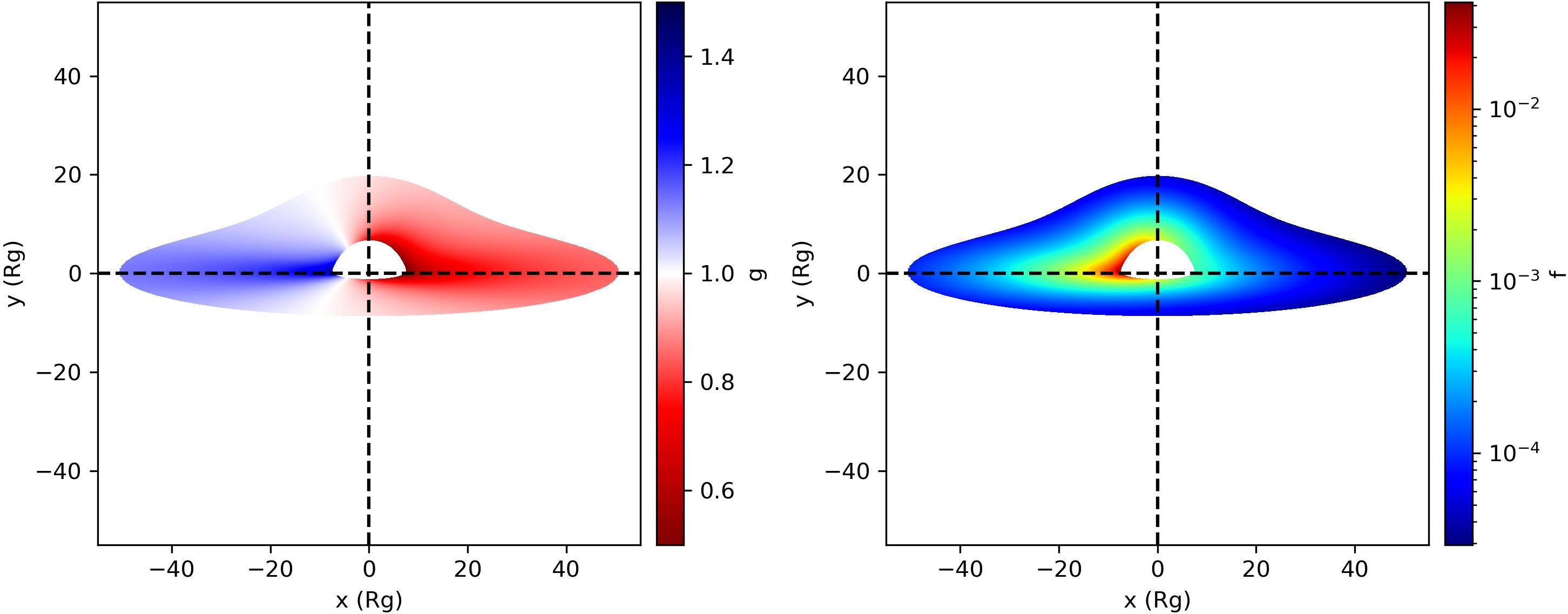}
\hfill
\includegraphics[width=0.28\textwidth]{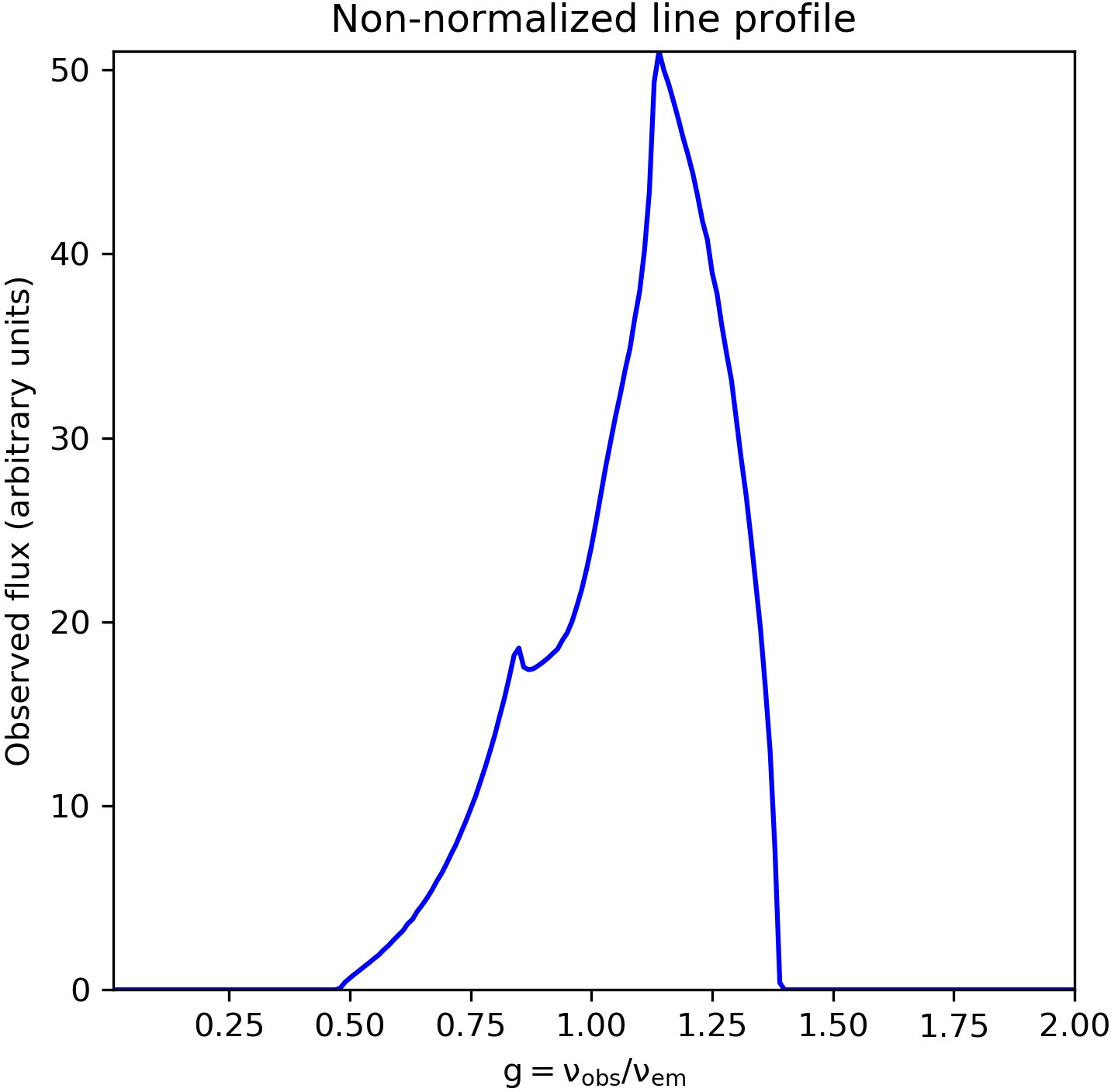}\\
\includegraphics[width=0.71\textwidth]{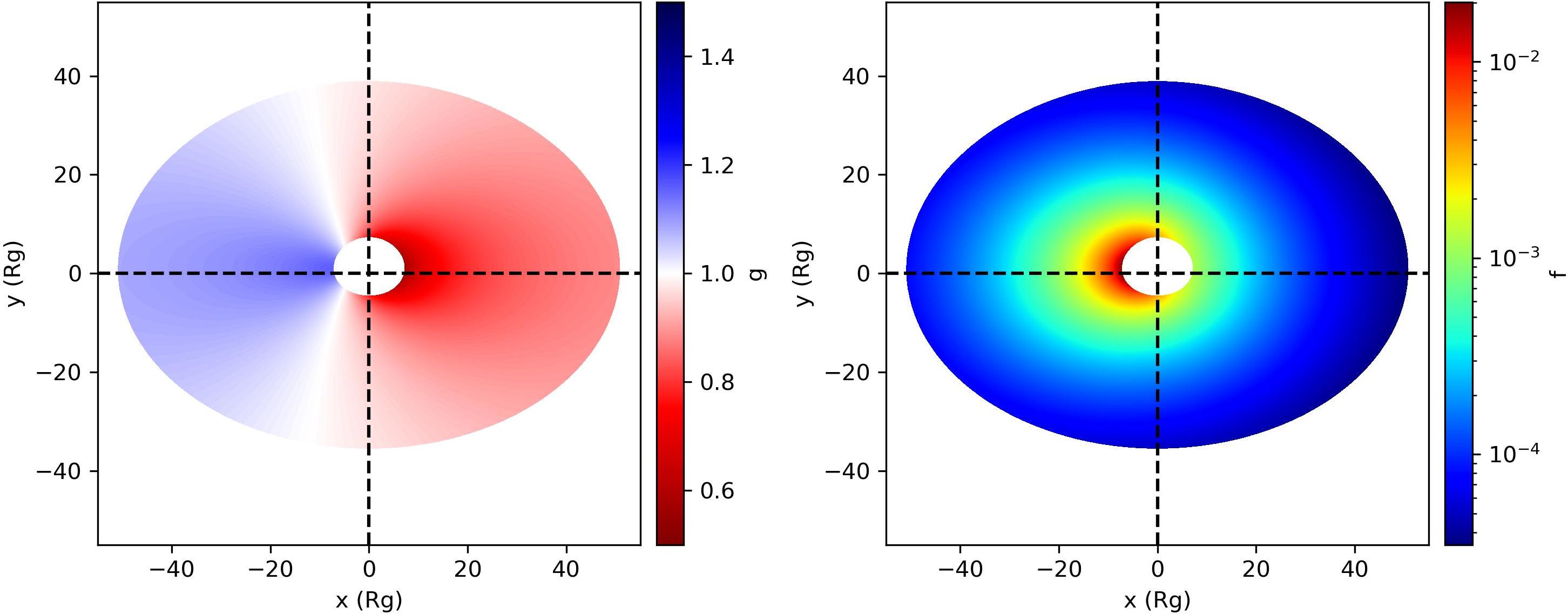}
\hfill
\includegraphics[width=0.28\textwidth]{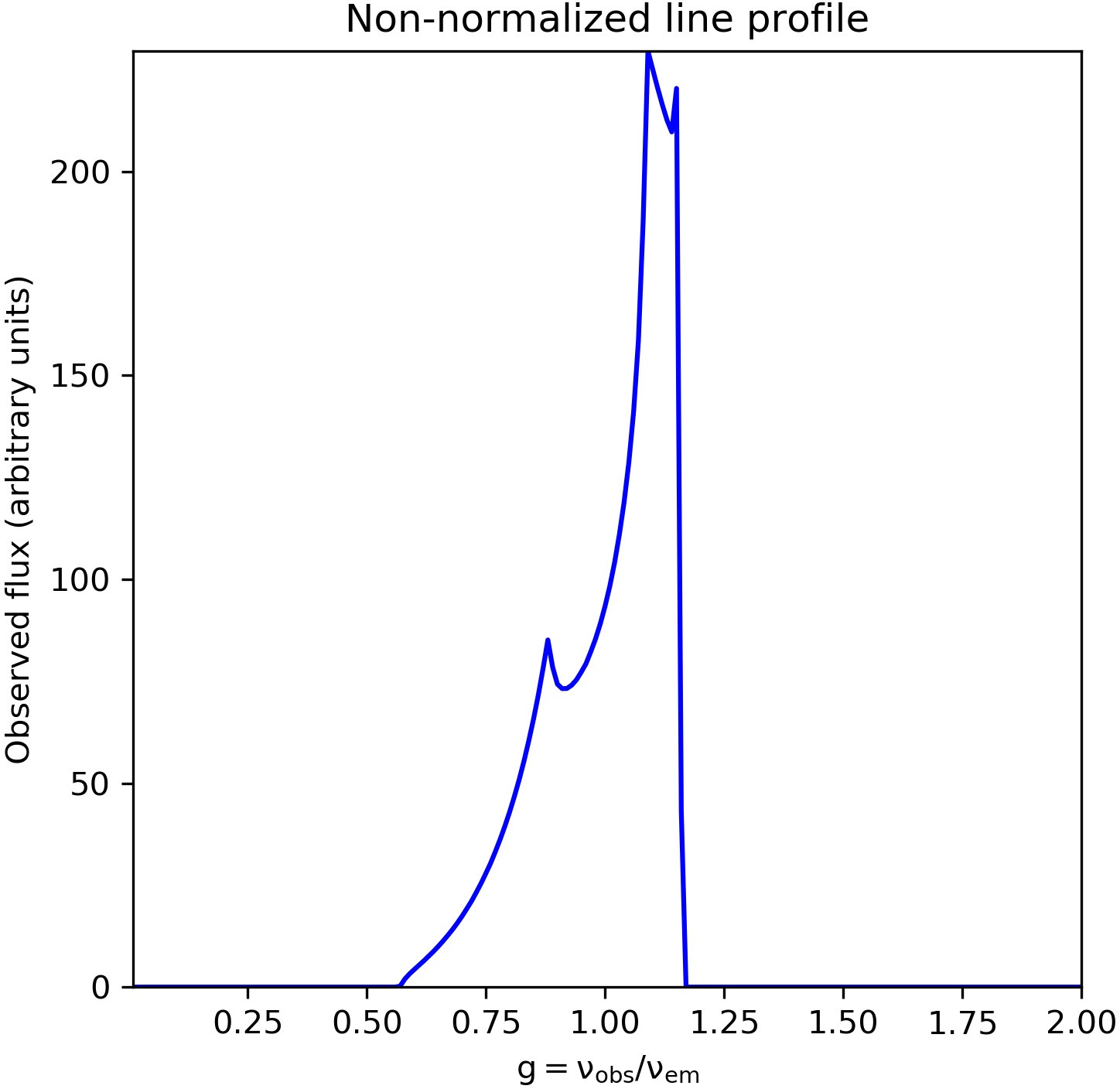}
\caption{Simulated images of the relativistic accretion disk in Kerr metric 
around a SMBH, colored according to energy shift $g$ (left) and observed flux (middle), as well as the corresponding simulated non-normalized line profile (right). Panels, from top to bottom, correspond to disk models 1, 2, 3 and 4, respectively (see \S\ref{sec:disks} for the particular values of disk parameters).}
\label{fig:disks}
\end{figure*}

\section{Modeling the optical and X-ray counterpart emission of SMBHB}

It is well known that the Fe K$\alpha$ line is produced from a very compact 
region around a SMBH, due to fluorescent emission \cite{fabi89,tana95,iwas96,fabi00}. This is the 
reason why the Fe K$\alpha$ line profile could be a powerful diagnostic tool for studying different physical and geometrical properties of such regions \citep[see e.g.][]{jova08,jova09,jova12,pop12},
as well as the masses and spins of SMBHs \citep[see e.g.][]{bren06,jova11,fero12,rey14}. Nowadays,
it is proven that X-ray reflection spectroscopy (based on the studies of the observed broad Fe K$\alpha$ line profiles)
became a powerful technique for robust black hole spin measurements across a wide range of
black hole masses (for more details, see the review in \cite{rey14}, and 
references therein). This technique could be applied in the range of the stellar-mass black holes 
in the X-ray binaries to the SMBHs in the Active Galactic Nuclei
(AGNs), as well as for the reverberation of the relativistically broadened Fe K$\alpha$ line, which
is already detected in the observed X-ray spectra of some AGNs \citep[e.g.][]{rey14,reyn99}.

Let us assume a realistic set of orbital elements and mass ratios for such a SMBHB and use them to simulate the corresponding composite H$\beta$ spectral line in the optical band and the Fe K$\alpha$ line in the X-ray band. The H$\beta$ line is assumed to come from two BLRs that follow the dynamics of component motions and an additional circumbinary BLR that surrounds the SMBHB system \citep[see][]{popo21}. Regarding the X-ray band, we studied the composite Fe K$\alpha$ line using the ray tracing method in Kerr metric, assuming that both accretion disks around primary and secondary give a significant contribution to the total Fe K$\alpha$ line emission of such a SMBHB \citep[see][and references therein]{jova20}.

\subsection{Optical variability of sub-pc SMBHB}
\label{sec:opt_var}

We examine the optical spectrum of such objects by applying the model, which simulates the binary system of supermassive black holes in AGN cores. Here we give a brief description of the model we used, and for more detailed information, see \cite{popo21}.

Basically, the model assumes two interacting black holes orbiting around a common mass center in Keplerian orbits. Their paths determine the plane of rotation, which could be inclined to the observer at an arbitrary angle. Both components of this system have their respective accretion disks, which are parallel to the orbiting plane. In the compact system we are dealing with, mutual tidal interaction can be important, particularly in medium and highly elliptical orbits. We included it by computing the mutual perturbation of the temperature profile of the disks and opposed BH. This perturbation induce additional variation of the continuum output emission during the full orbiting phase of the system.

Additionally, dusty material in the surrounding region of the disk \citep[see][]{chen89,czerny11,song21,naddaf22}, form
the potential ionization regions, where a discrete spectrum is generated by ionization processes. Those regions (called Broad Line Regions, or BLRs) are gravitationally bound to their BHs and orbit together, following the same path. Since we are here discussing the compact binary system, it is convenient to assume that BLRs have some kind of size limitation intruded by the tidal and inertial forces of both components. Therefore, we assumed that BLR size is determined by the Roche radius defined for such a binary system:

\begin{equation}
r_i=A(t)\frac{0.49q^{2/3}}{0.6q^{2/3}+\ln(1+q^{1/3})}
\label{eq:roche}
\end{equation}
where $A(t)$ is current component distance and $q$ mass ratio.
In this case, one can expect that the intensity of the broad line emission is proportional to the size of this region. Therefore, the contributions of BLR1 and BLR2 to the total broad line intensity are proportional to their dimensions. Additionally, we assumed that material in these regions is virialized, which allows us to examine velocity distribution and consequently line widths. In the proposed configuration, we included a circum-binary region surrounding both components and their BLRs. This area is populated with the same ionized material, but it is stationary, consequently contributing to the total line spectrum without causing the line shifts.

\subsection{Gaussian line profiles for optical band}

In case of lines in the optical part of spectrum, geometry and mass distribution is of particular importance. Here we consider that all mass is mainly located in the orbiting plane of the BH components, which we call flattened geometry \cite{ga09}. This imply that accretion disks and BLRs from both BHs are in same plane, defined with the inclination angle $i$ toward the observer. Additionally, for simplicity reasons we consider a cBLR mass distribution to be also coplanar with the orbital plane, although different configurations can be considered.

The BLR can have a complex structure (disk-like structure, outflows and inflows, etc.) that may result in complex broad line shapes and line profile asymmetry. However, most of the broad lines have Gaussian-like profiles where the line width is affected by the orbital motions of the emitting clouds around the supermassive black hole and can be used for black hole mass estimation \citep[see a review in][]{pop20}. Therefore, in our model, which considers the full binary dynamical effect, we assumed that all three broad line regions are emitting Gaussian-like profiles (we take a pure Gaussian) where the velocity dispersion of the Gaussian is connected with the mass of the central supermassive black hole component. In the case of cBLR, we assume that velocity dispersion is caused by the sum of the masses of both components.

As a result of such distribution we expect that double-peaked spectral lines could be produced instead of Gaussion profile. From the point of the observations, this is desirable profile, since it offers an easy way for detection of those objects. 
However, we consider a complex system containing different regions with different thermodynamical and kinematical conditions. Total emission is a superposition of the contributions from all regions.
Therefore, combined line deviate from double-peaked profile and it is mostly similar to Gaussian, with possible asymmetric deviation in the wings or additional small superposed peak.

Since we are dealing with very compact system, we consider that BLRs are clearly separated in the Roche lobes of components 1 and 2, each with $\mathrm{H_\beta}$ line emission contribution of $I_1(\lambda)$ and $I_2(\lambda)$, respectively. Combined line is calculated by superposition of contribution from each component as:
\begin{equation}
I_{\mathrm{dyn}}(\lambda)=I_1(\lambda)+I_2(\lambda),
\label{eq:I_tot}
\end{equation}
Explicit form of $I_i(\lambda)$ with proposed Gaussian line profile is:
\begin{equation}
I_{i}(\lambda)=I_{i}(\lambda_0)\exp{\left[-\left(\frac{\lambda-\lambda_{0}\cdot (1+z^{i}_{\mathrm{dopp}})}{\sqrt{2}\sigma_{i}} \right)^2 \right]}\cos(i),
\label{eq:line}
\end{equation}
where $\lambda_0$ is the laboratory wavelength for H$\beta$, $z^i_{\mathrm{dopp}}$ is the Doppler correction for radial component velocities, and $\sigma_{i}$ is the dispersion of the velocity determined by the SMBH mass and size of the BLR (see Eqs. (\ref{eq:sigma_od_v}) and (\ref{eq:v_blr}) and the following text). Additionally, contribution from the cBLR, has to be accounted, then we have the total line profile as:
\begin{equation}
I_{\mathrm{tot}}(\lambda)=I_\mathrm{dyn}(\lambda)+I_\mathrm{cBLR}(\lambda),
\label{eq:I_tot1}
\end{equation}
with cBLR component contributing the following Gaussian line profile:
\begin{equation}
I_\mathrm{cBLR}(\lambda)=I_\mathrm{cBLR}(\lambda_0)\exp{\left[-\left(\frac{\lambda-\lambda_{0}}{\sqrt{2}\sigma_\mathrm{cBLR}} \right)^2 \right]}\cos(i),
\label{eq:line1}
\end{equation}
where $\sigma_\mathrm{cBLR}$ is the velocity dispersion determined by total mass which includes both components and dimension of the cBLR.

To compute line profiles given by Eqs. (\ref{eq:line}) and (\ref{eq:line1}) we need to estimate line intensities of BLR $I_i(\lambda_0)$ and circum-binary region $I_{cBLR}(\lambda_0)$. We used empirical relation connecting black hole mass and $\mathrm{H_{\beta}}$ line luminosity for single-epoch, as given by \cite{bonta20}:
\begin{equation}
\begin{split}
\log\mu(H_\beta)&=6.975+0.566\left[\log\ L(H_\beta) - 41.857\right] \\
& + 1.757\left[\log\sigma_M(H_\beta)-3.293\right],
\label{eq:mu_Hb_empirical}
\end{split}
\end{equation}
where $\mu$ is related to the BH mass as $\log M = \log f + \log\mu$.

Line luminosities and intensities are connected as:
\begin{equation}
I_{i}(\lambda_0)={\lambda L(\mathrm{H}\beta)\over{\sqrt{2\pi}\sigma_i}},
\label{eq:I_lambda_0}
\end{equation}
where velocity dispersion $\sigma_i$ is expressed with the BLR velocity $v_\mathrm{BLR}$ as:
\begin{equation}
\sigma_i=\lambda_{\mathrm{H}\beta}\frac{v_{\mathrm{BLR}}(m_i)}{c}.
\label{eq:sigma_od_v}
\end{equation}
Assuming that the BLR is mostly virialized, we can calculate the velocity as:

\begin{equation}
v_{\mathrm{BLR}}(m_i)=\sqrt{\frac{Gm_i}{f_VR_{\mathrm{BLR}}}},
\label{eq:v_blr}
\end{equation}
where $R_{\mathrm{BLR}}$ is BLR size, $m_i$ SMBH mass, $G$ gravitational constant and $f_V$ is the virialization factor, defined by BLR geometry and inclination as $f_V=1/\sin^2(i)$ \citep[see][]{af19}. This simple form of $f_V$ corresponds to the case when Keplerian motion of the gas is dominant, and for more general form of $f_V$ see \citet{collin06}. In the same way, $v_{\mathrm{cBLR}}$ can be calculated, with the mass as ($m_1+m_2$) instead of the mass of one component ($m_i$) in Eq. (\ref{eq:v_blr}).

As we can see in the previous equations, velocity of the gas in the BLRs directly depends on the SMBH mass and dynamical parameters. Therefore, the line profiles will depend on the same parameters, as well as line maximal intensity.
Several authors pointed out that the broad line profiles could be distorted by gravitational redshift \citep[see e.g.][]{netz77,netz90}. However, since BLR is at larger distance from the SMBH, both transverse and gravitational redshifts are negligible. Thus, we can estimate the gravitational redshift in BLR as \citep[see e.g.,][]{jon16}:
\begin{equation}
\Delta\lambda_\mathrm{g}^i=\frac{Gm_i}{cR_{\mathrm{BLR}}}.
\end{equation}

The estimation gives a gravitational redshifts around 50 km s$^{-1}$ for the BLR size of 10 light days. Also, the transverse redshift is a half of the gravitational one (see Eq. (23) in \citet{puns20}), and both are significantly smaller that the dynamical shift which is the order of 1000 km s$^{-1}$. Therefore, we did not take them into account.

\subsection{X-ray variability of sub-pc SMBHB}

We study the potential observational signatures of SMBHBs in the Fe
K$\alpha$ line profiles emitted from the relativistic accretion disks around
their SMBH components. In our numerical simulations of disk emission, we used
the ray-tracing method in the Kerr metric. Also, we take into account only those
photon trajectories which reach the observer's sky plane
\citep{fant97,cade98,jova08,jova09,jova12}. The disk emissivity
$\varepsilon \left( {r} \right)$ is modeled with the power law:
$\varepsilon\left({r}\right)=\varepsilon_{0}\cdot{r^p}$, where
$\varepsilon_{0}$ is an emissivity constant, and $p$ is an emissivity index
\cite{jova12}.

\subsection{Simulated line profiles in the X-ray band}

We can define the energy shift $g$ of photon which represents the ratio between $E_{em}$ and $E_{obs}$ (emitted and observed photon energy, respectively), or using corresponding photon wavelengths $\lambda_{em}$) and $\lambda_{obs}$ (emitted and observed photon wavelength, respectively), or using the usual redshift in wavelength $z$ \citep[
see e.g.][and references therein]{jova16}:
\begin{equation}
\label{eqn:shift}
g=\dfrac{E_{obs}}{E_{em}}=\dfrac{\lambda_{em}}{\lambda_{obs}}=\dfrac{1}{1+z}.
\end{equation}

The observed flux at each observed energy over the whole disk image one can obtain by integration the corresponding simulated line profile \citep{jova12}:
\begin{equation}
\label{eqn:line}
F_{obs} \left( {E_{obs}} \right) = {\displaystyle\int\limits_{image}
{\varepsilon\left({r} \right)}} g^{4}\delta \left( {E_{obs} - gE_{0}}
\right)d\Xi .
\end{equation}

Under an assumption that the radial velocities of SMBHB components are non-relativistic (i.e. $V_{1,2}^{rad} \ll c$), the
corresponding Doppler shifts in wavelength ($z_{1,2}$) and energy ($g_{1,2}$) due to such radial velocities are then given by:
\begin{equation}
\label{eqn:dshift}
z_{1,2}\approx
\dfrac{V_{1,2}^{rad}}{c},\quad g_{1,2}=\dfrac{1}{1+z_{1,2}} .
\end{equation}
We can also obtain the total redshift factor $g_{tot}$ from Eqs. (\ref{eqn:shift}) and (\ref{eqn:dshift}):
\begin{equation}
\label{eqn:gtot}
g_{tot}=\dfrac{1}{1+z+z_{1,2}}=\dfrac{1}{\dfrac{1}{g}+\dfrac{1}{g_{1,2}}-1}.
\end{equation}
It represents the net effect of both relativistic effects and radial velocities of the components. Using  expression (\ref{eqn:line}), we studied the influence of Doppler shifts on the observed disk emission, but using the total redshift factor $g_{tot}$ instead of $g$.

\begin{figure*}[ht!]
\centering
\includegraphics[width=1\textwidth]{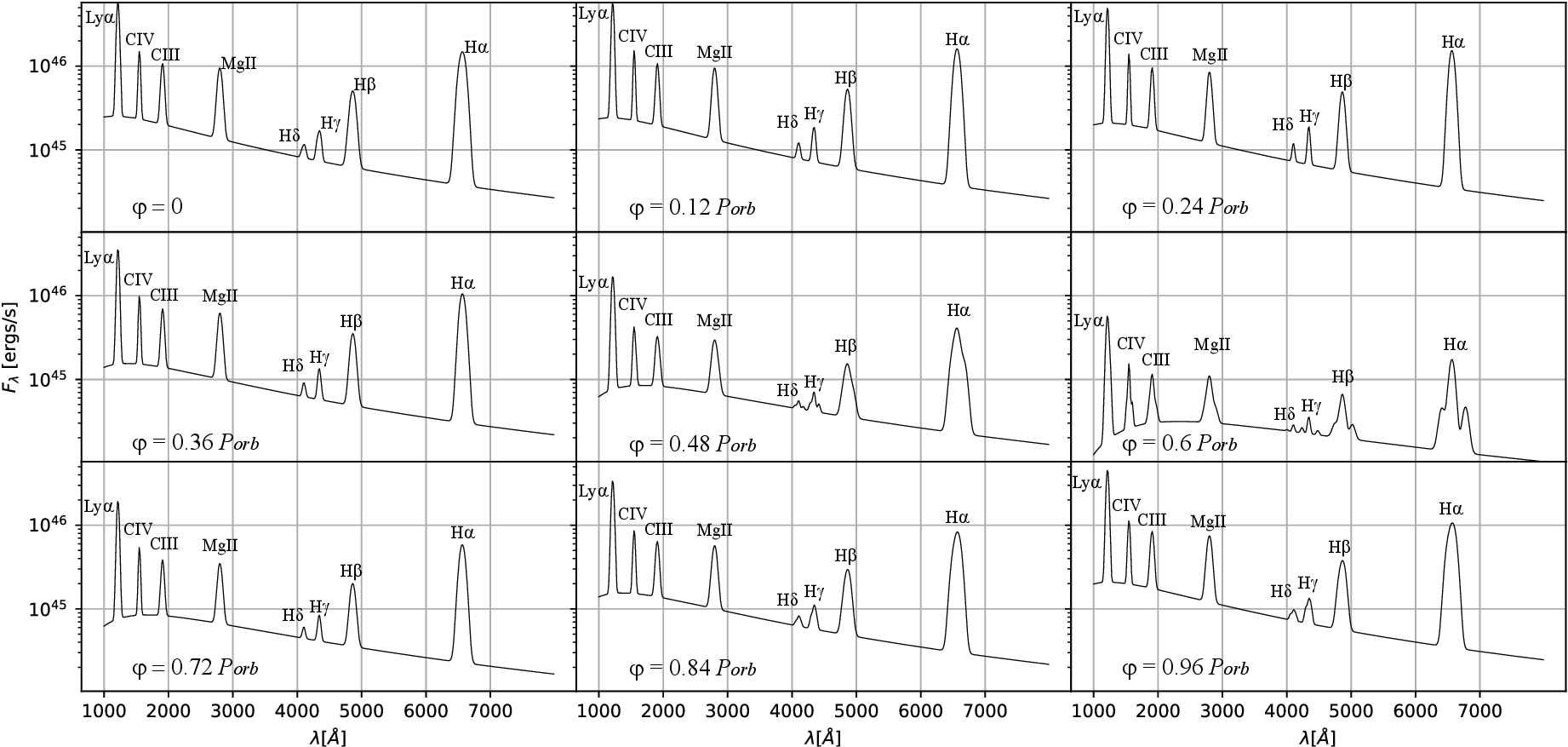}
\caption{Variation of the broad optical spectrum (spectrum lines + continuum) of the discussed binary system for nine different orbital phases of the binary system. We note that the line contribution from the NLR is not accounted for. Parameters used for generating this Figure are given in Table \ref{tab:param_set}.}
\label{fig:broad_optical}
\end{figure*}

\begin{figure*}[ht!]
\centering
\includegraphics[width=1\textwidth]{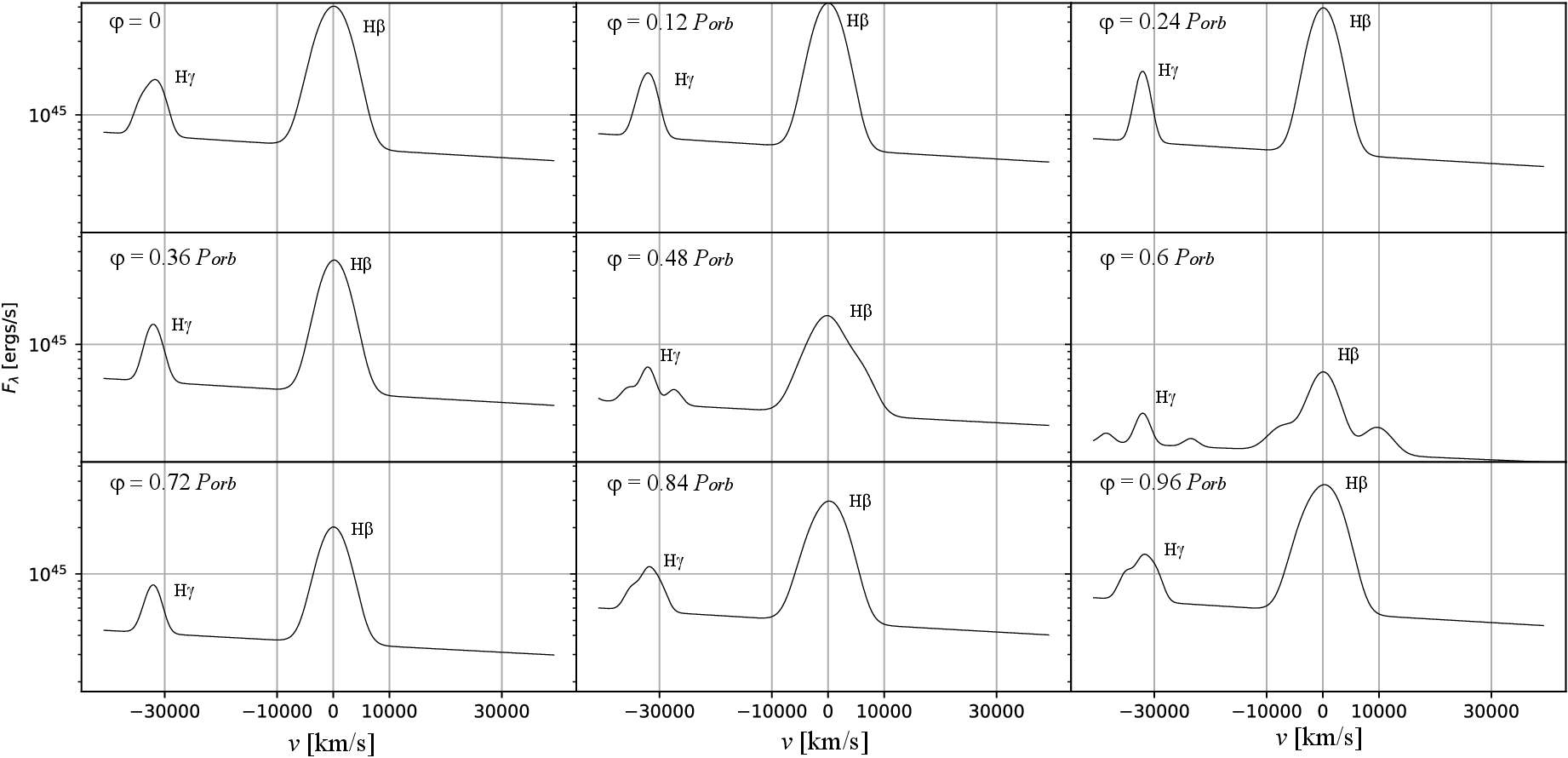}
\caption{Same as in Fig. \ref{fig:broad_optical}, but for variability of the H$\beta$ line.}
\label{fig:HB_line_continuum}
\end{figure*}

\begin{figure*}[ht!]
\centering
\includegraphics[width=0.49\textwidth]{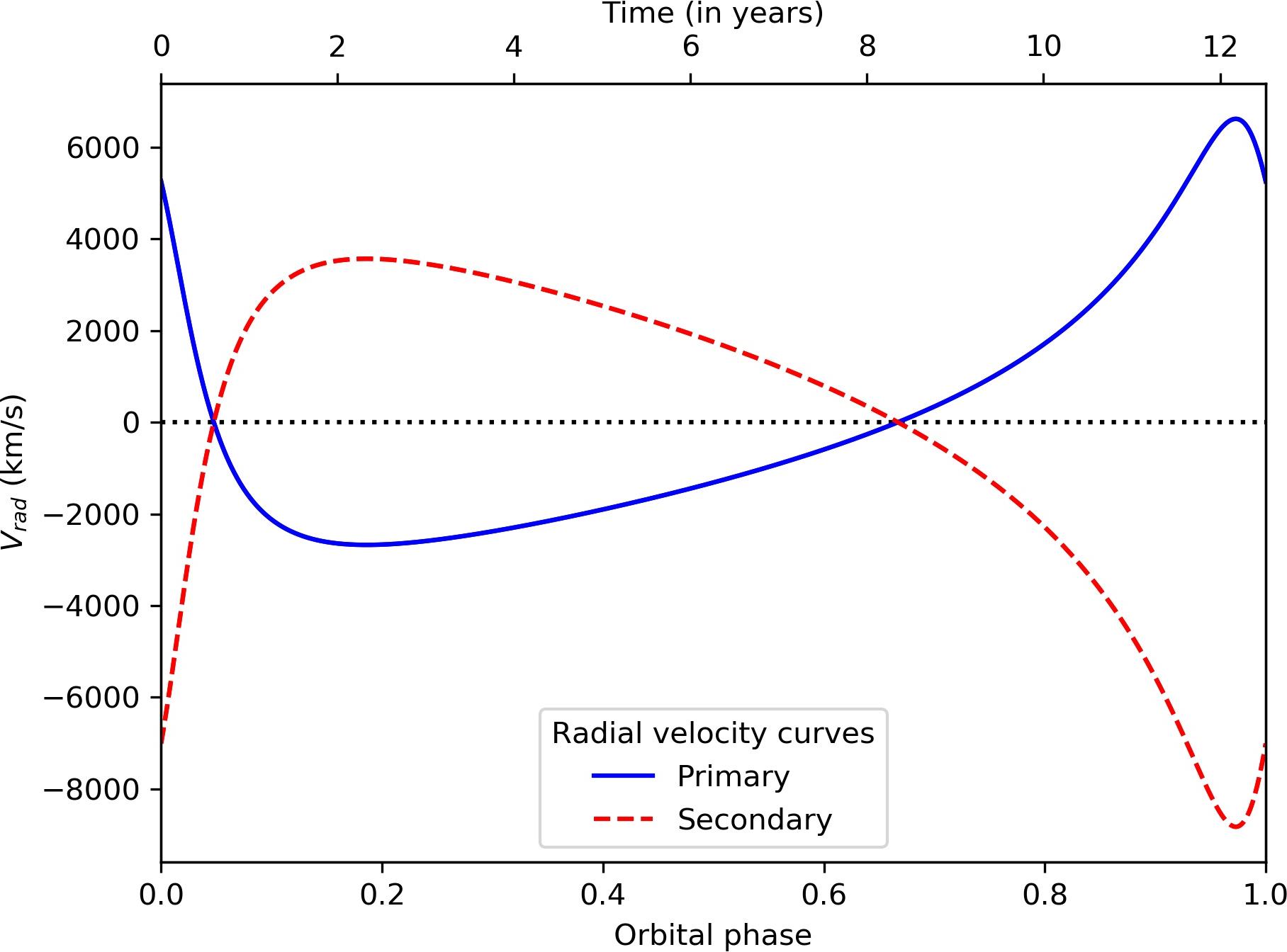}
\hfill
\includegraphics[width=0.49\textwidth]{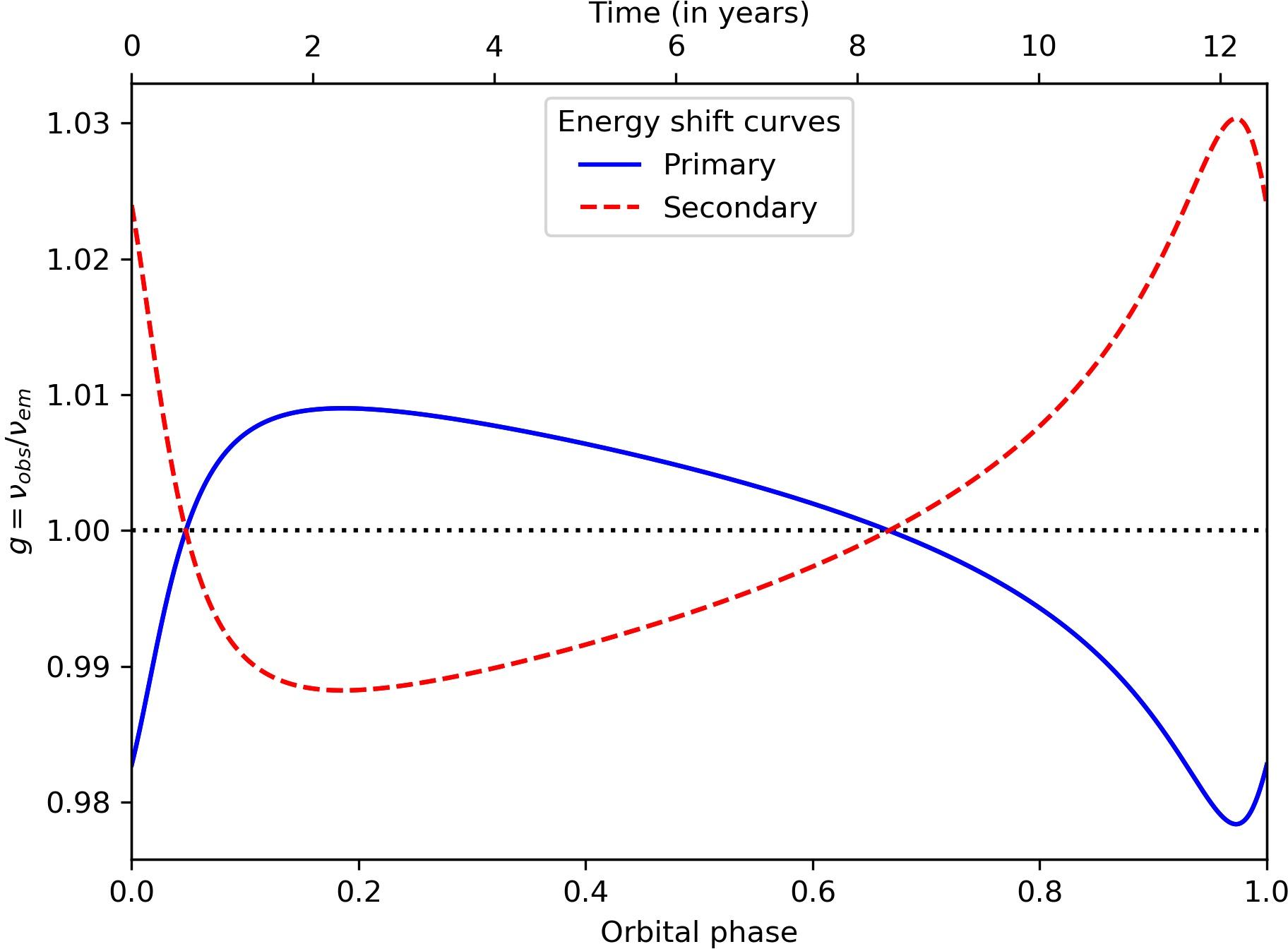}
\caption{Radial velocities (left) and redshift factors (right) of the components in a SMBHB with mass ratio q = 0.75, (see \S\ref{sec:smbhb} for the particular values of orbit parameters).}
\label{fig:vrad}
\end{figure*}

\begin{figure*}[ht!]
\centering
\includegraphics[width=0.48\textwidth]{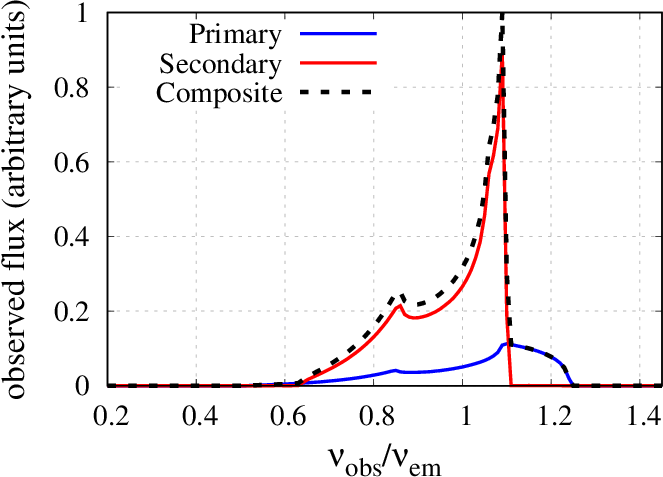}
\hfill
\includegraphics[width=0.48\textwidth]{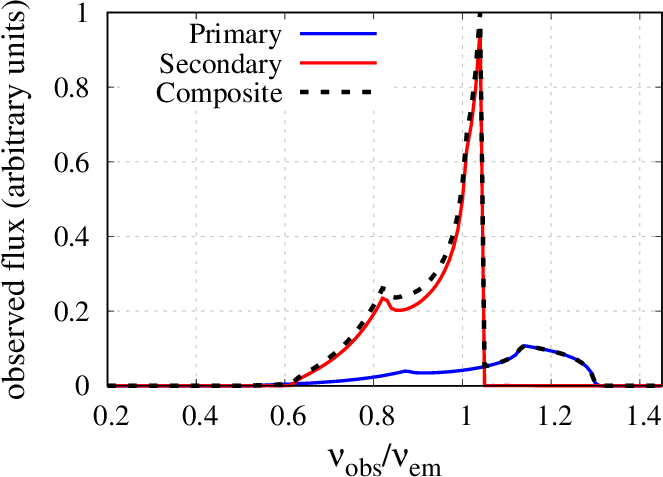}\\
\vspace{0.2cm}
\includegraphics[width=0.48\textwidth]{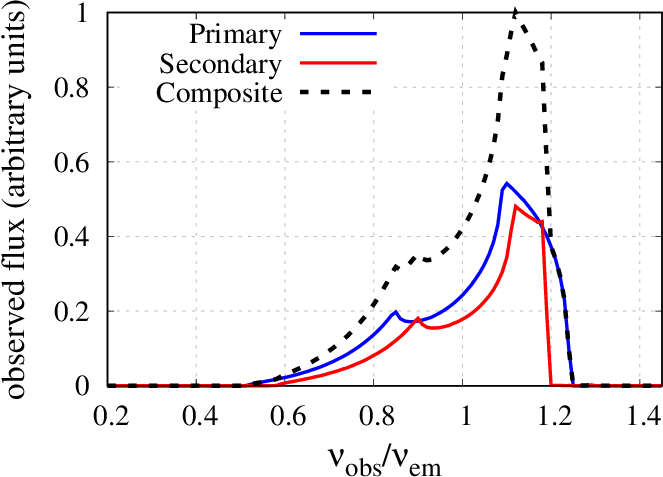}
\hfill
\includegraphics[width=0.48\textwidth]{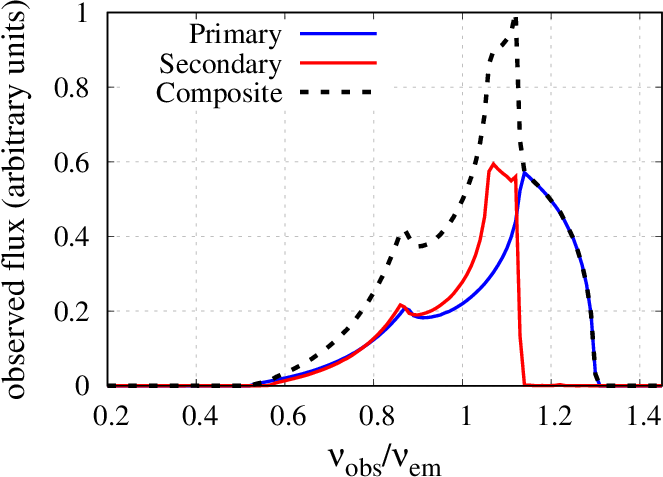}\\
\vspace{0.2cm}
\includegraphics[width=0.48\textwidth]{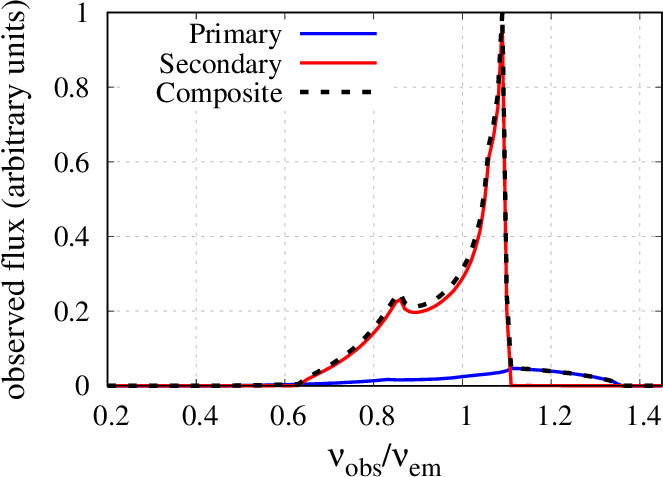}
\hfill
\includegraphics[width=0.48\textwidth]{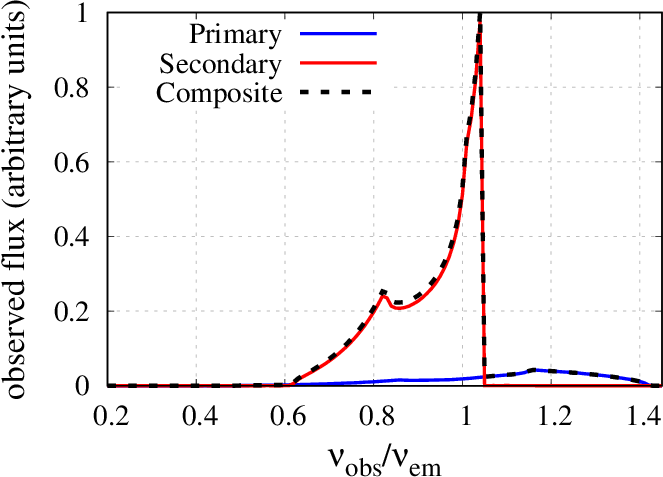}
\caption{\textit{Top:} The simulated Fe K$\alpha$ line profile from disk 1 around the primary (blue solid line), disk 2 around secondary (red solid line) and their  normalized composite profile (black dashed line), emitted during the pericenter (left) and apocenter orbital phase (right), respectively. \textit{Middle:} The same as in top panel, but for disk 1 around the primary and disk 4 around the secondary. \textit{Bottom:} The same as in top panel, but for disk 3 around the primary and disk 2 around the secondary.}
\label{fig:periapo}
\end{figure*}

\begin{figure*}[ht!]
\centering
\includegraphics[width=0.48\textwidth]{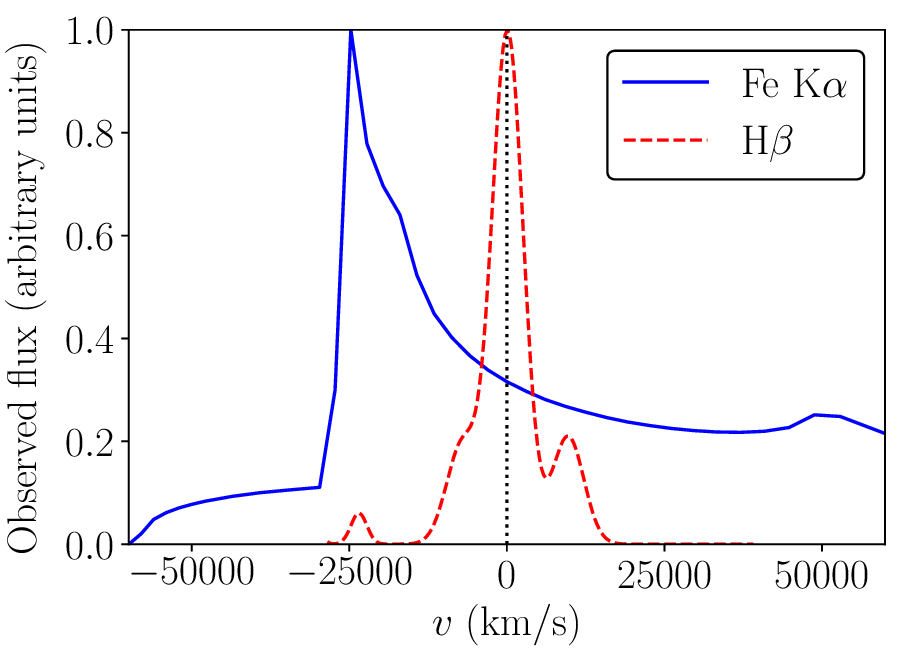}
\hfill
\includegraphics[width=0.48\textwidth]{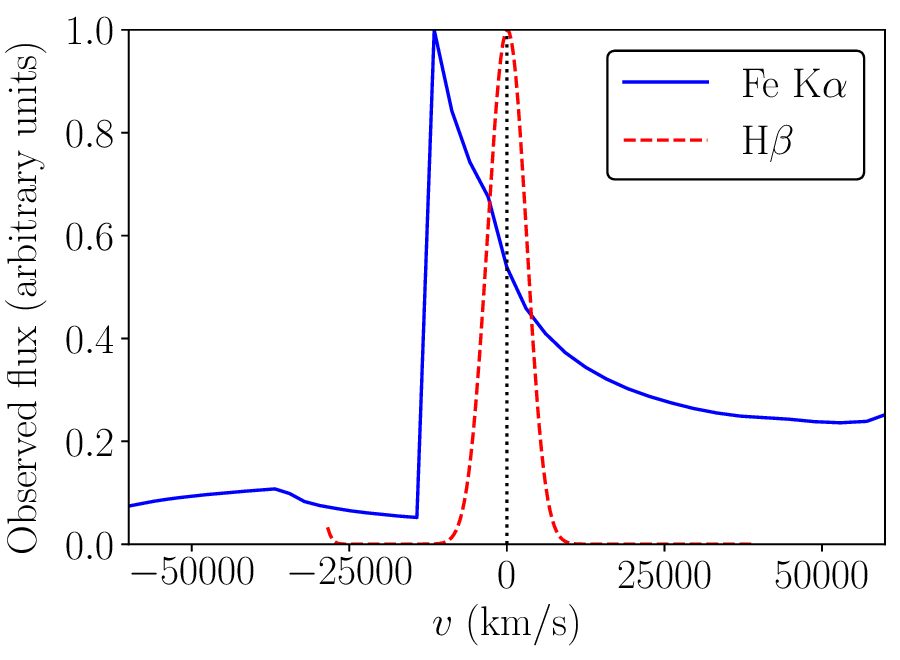}\\
\vspace{0.4cm}
\includegraphics[width=0.48\textwidth]{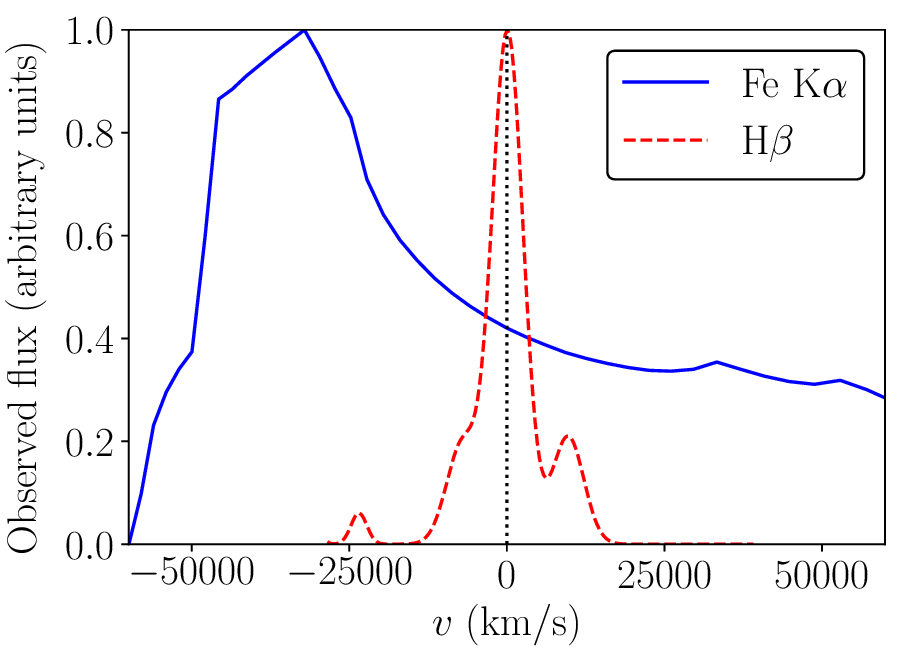}
\hfill
\includegraphics[width=0.48\textwidth]{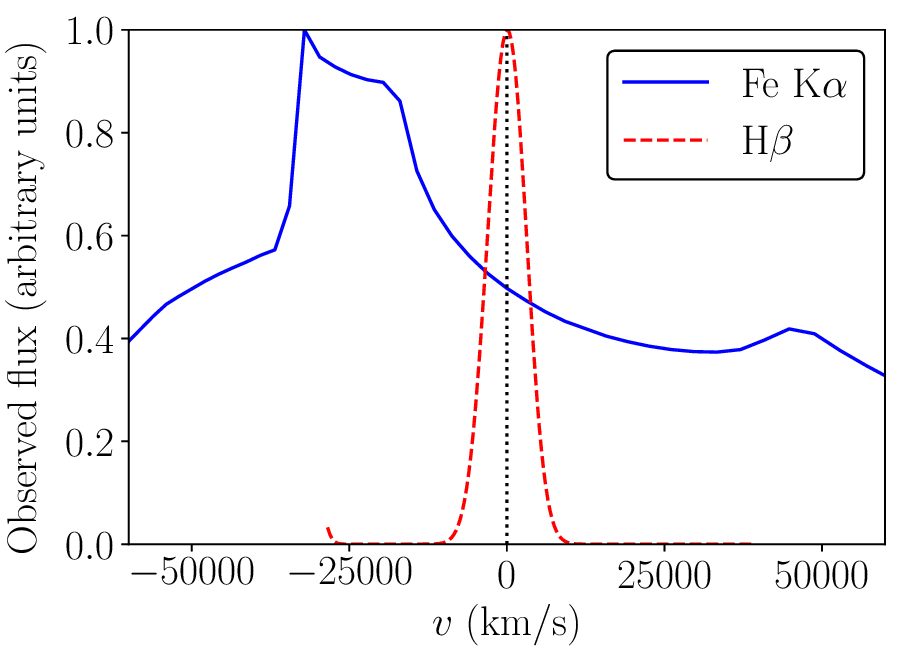}\\
\vspace{0.4cm}
\includegraphics[width=0.48\textwidth]{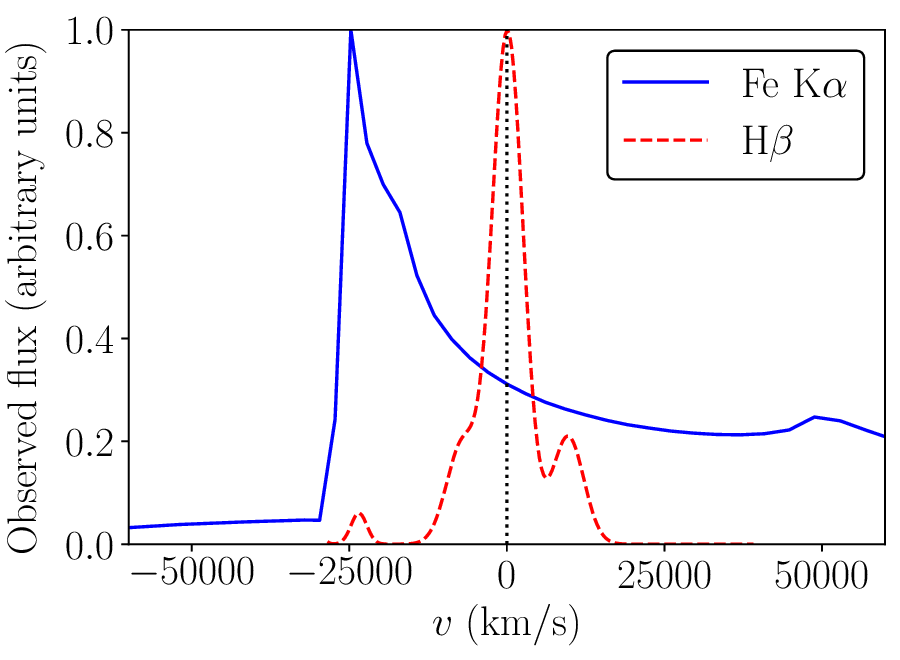}
\hfill
\includegraphics[width=0.48\textwidth]{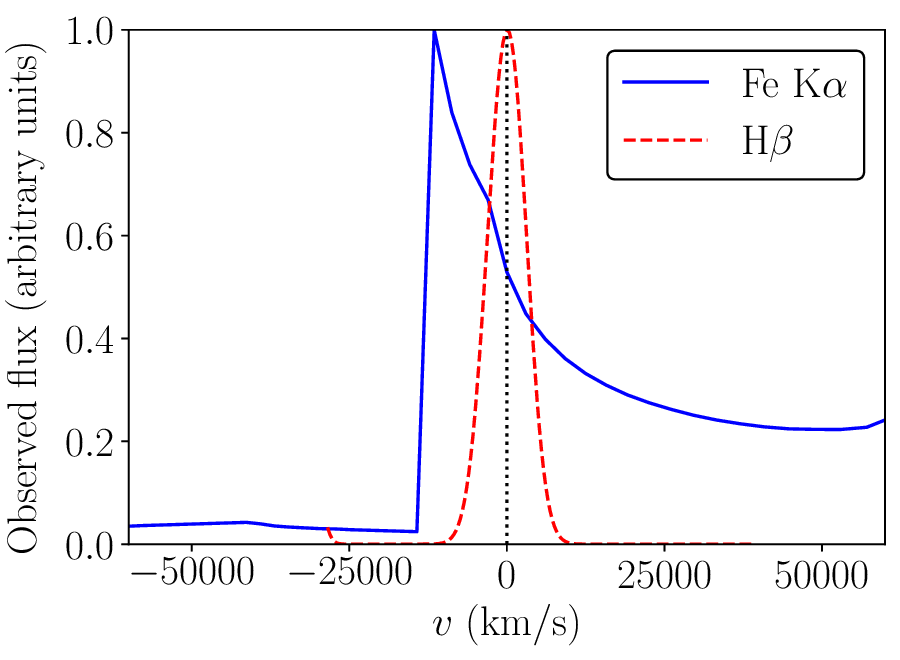}
\caption{\textit{Top:} Comparison between the entire normalized H$\beta$ line profile (red dashed line) and the central part of the normalized composite Fe K$\alpha$ line profile (blue solid line), in the case when the latter is emitted from disk 1 around the primary and disk 2 around secondary, during their pericenter (left) and apocenter orbital phase (right), respectively. \textit{Middle:} The same as in top panel, but for disk 1 around the primary and disk 4 around the secondary. \textit{Bottom:} The same as in top panel, but for disk 3 around the primary and disk 2 around the secondary.}
\label{fig:hbvsfa}
\end{figure*}

\begin{figure*}[ht!]
\centering
\includegraphics[width=0.49\textwidth]{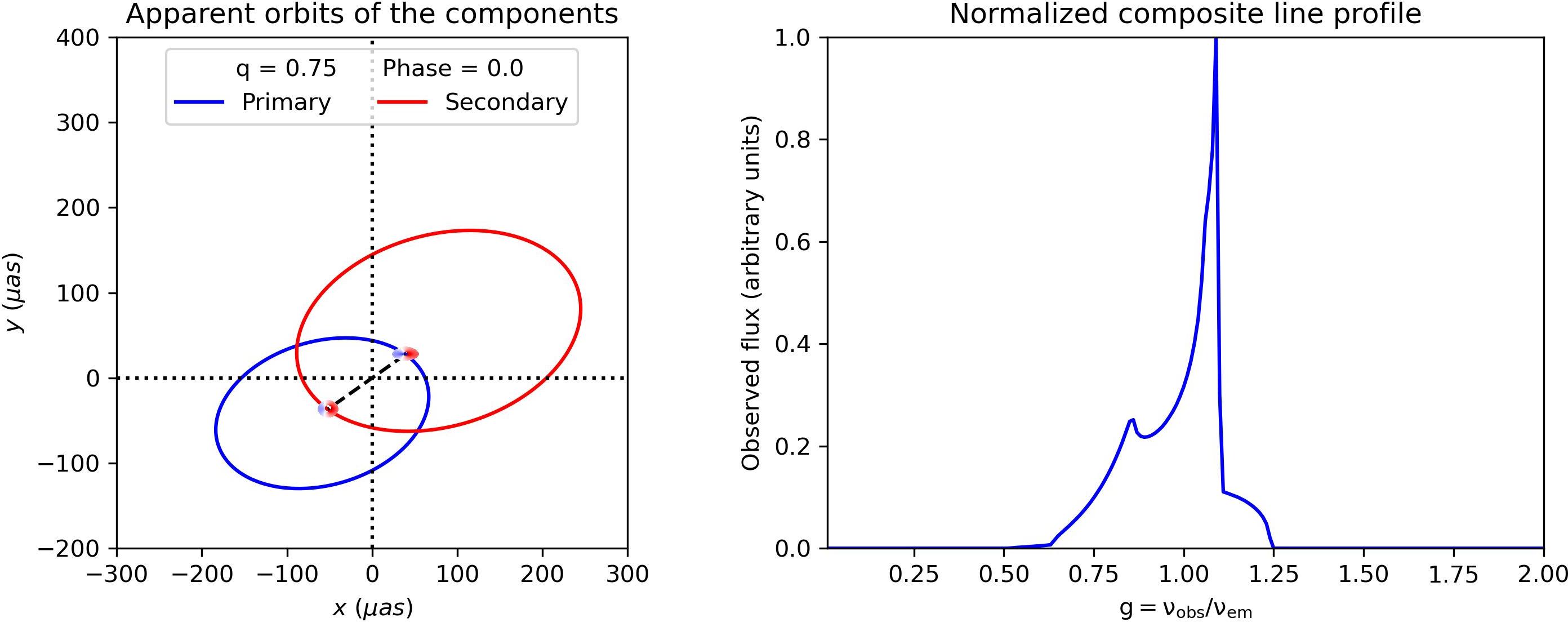}
\hfill
\includegraphics[width=0.49\textwidth]{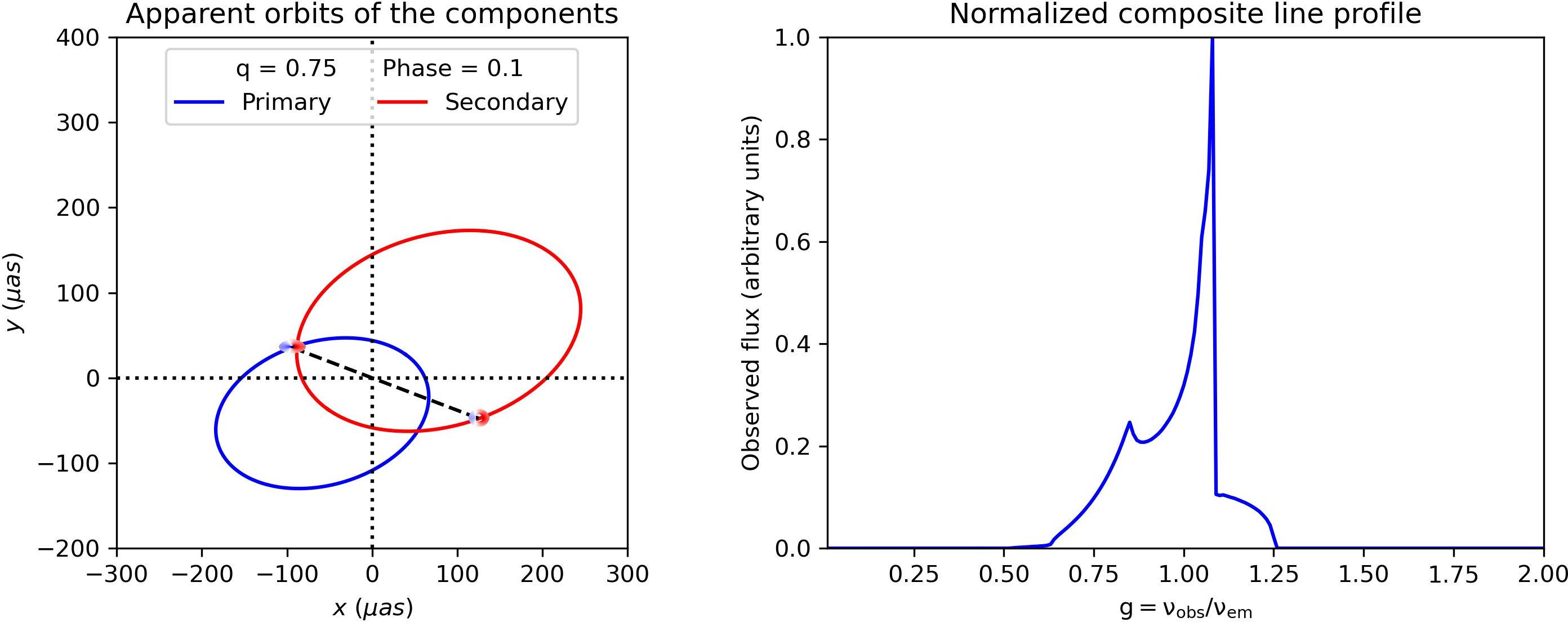}
\includegraphics[width=0.49\textwidth]{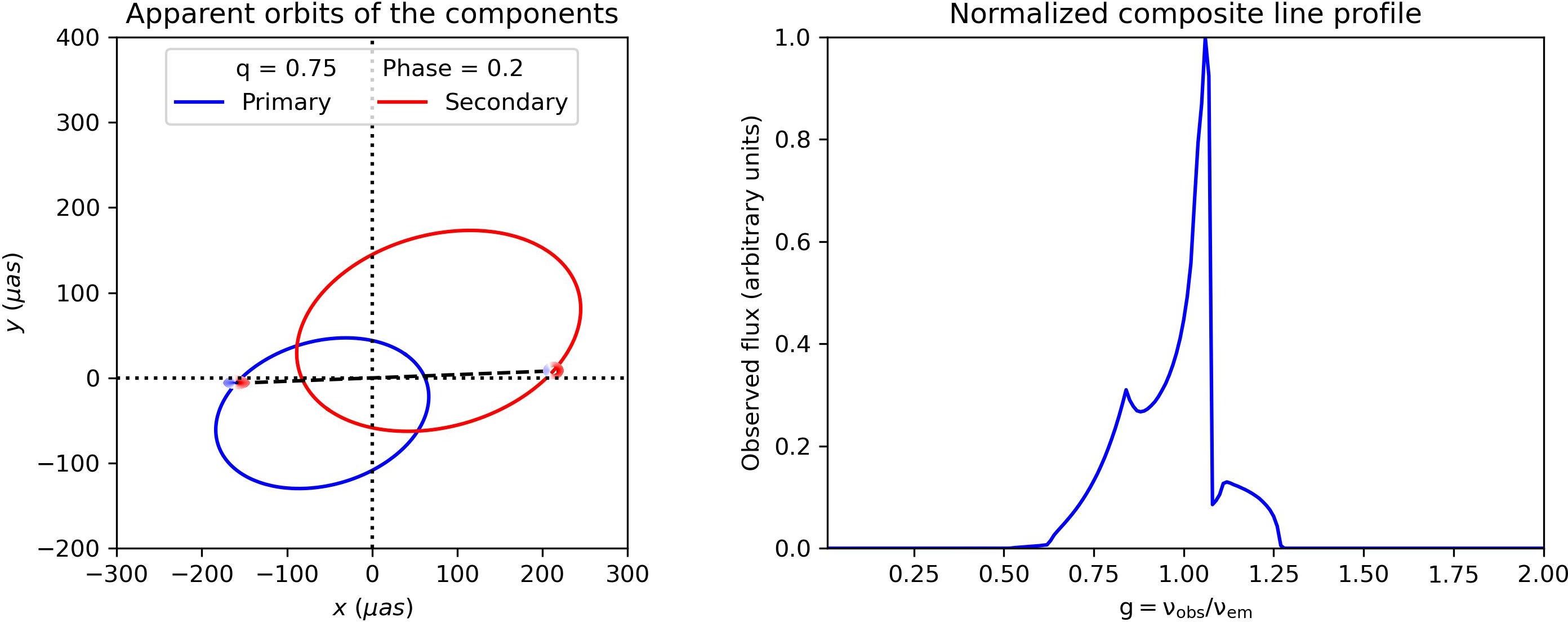}
\hfill
\includegraphics[width=0.49\textwidth]{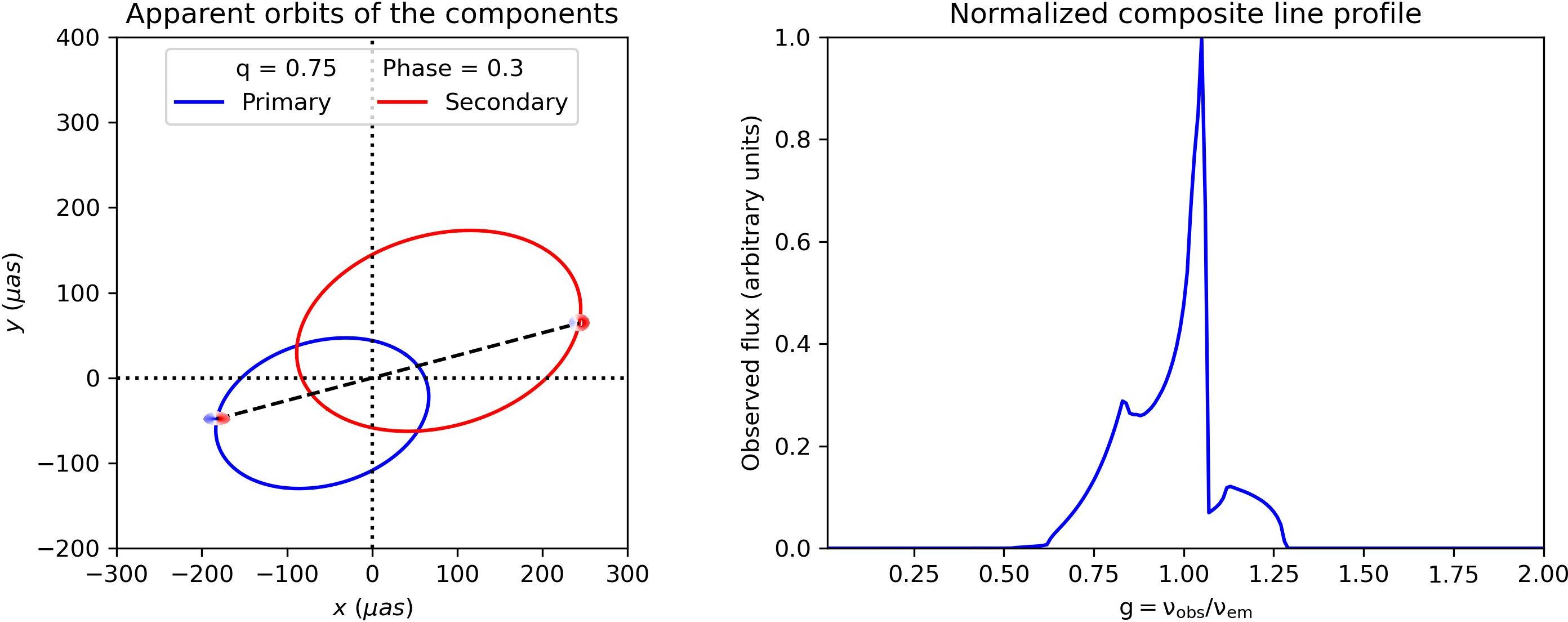}\\
\includegraphics[width=0.49\textwidth]{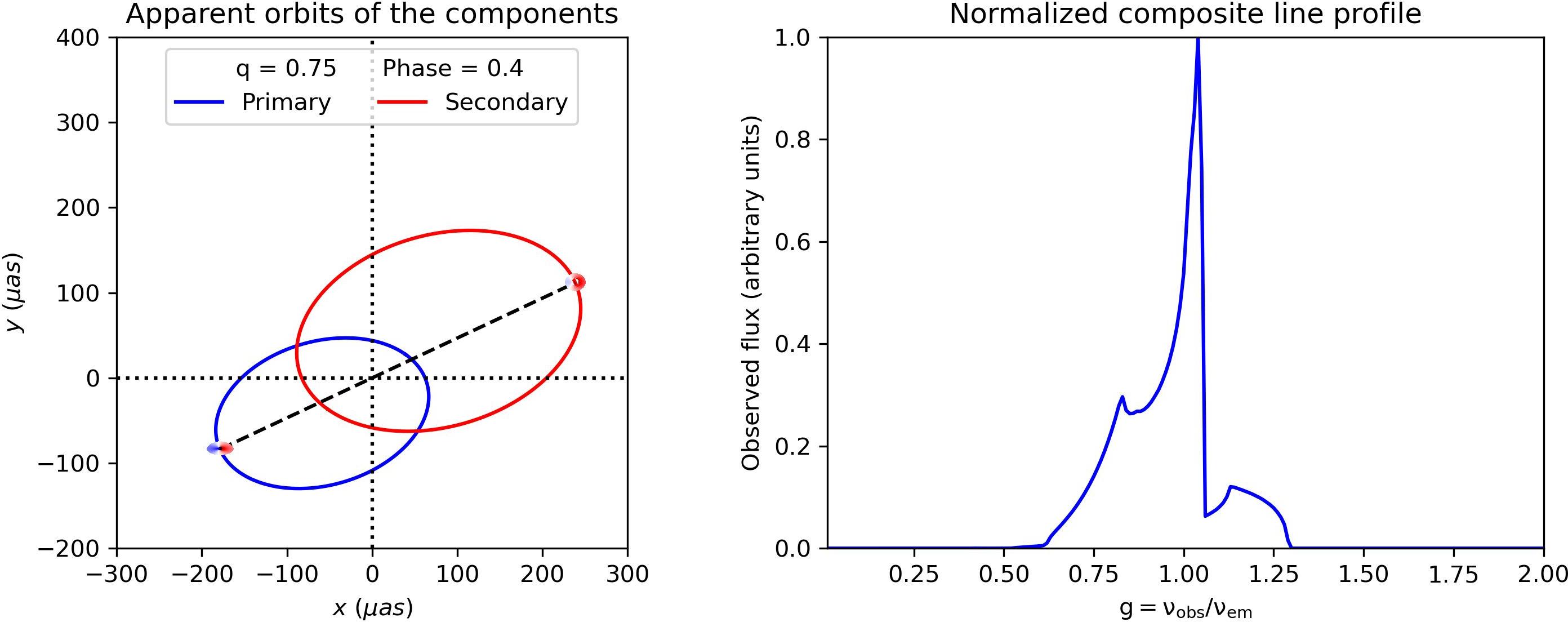}
\hfill
\includegraphics[width=0.49\textwidth]{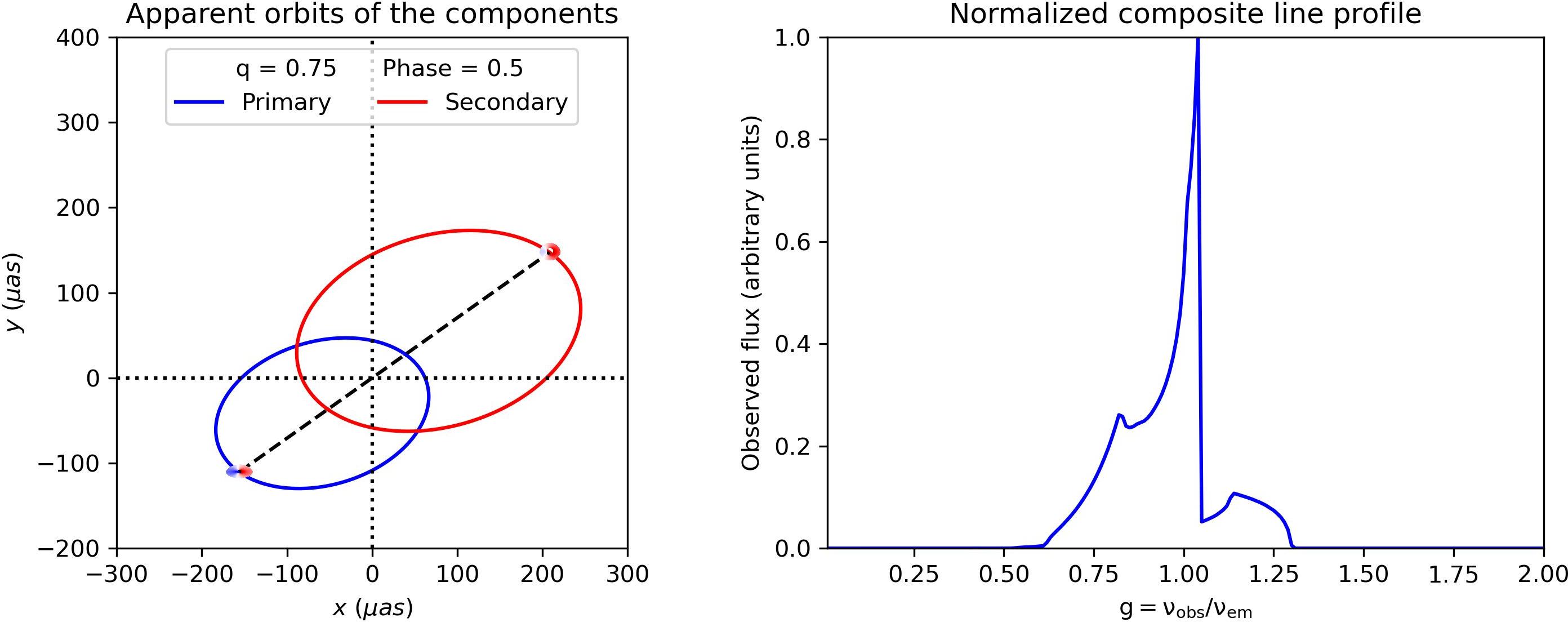}
\includegraphics[width=0.49\textwidth]{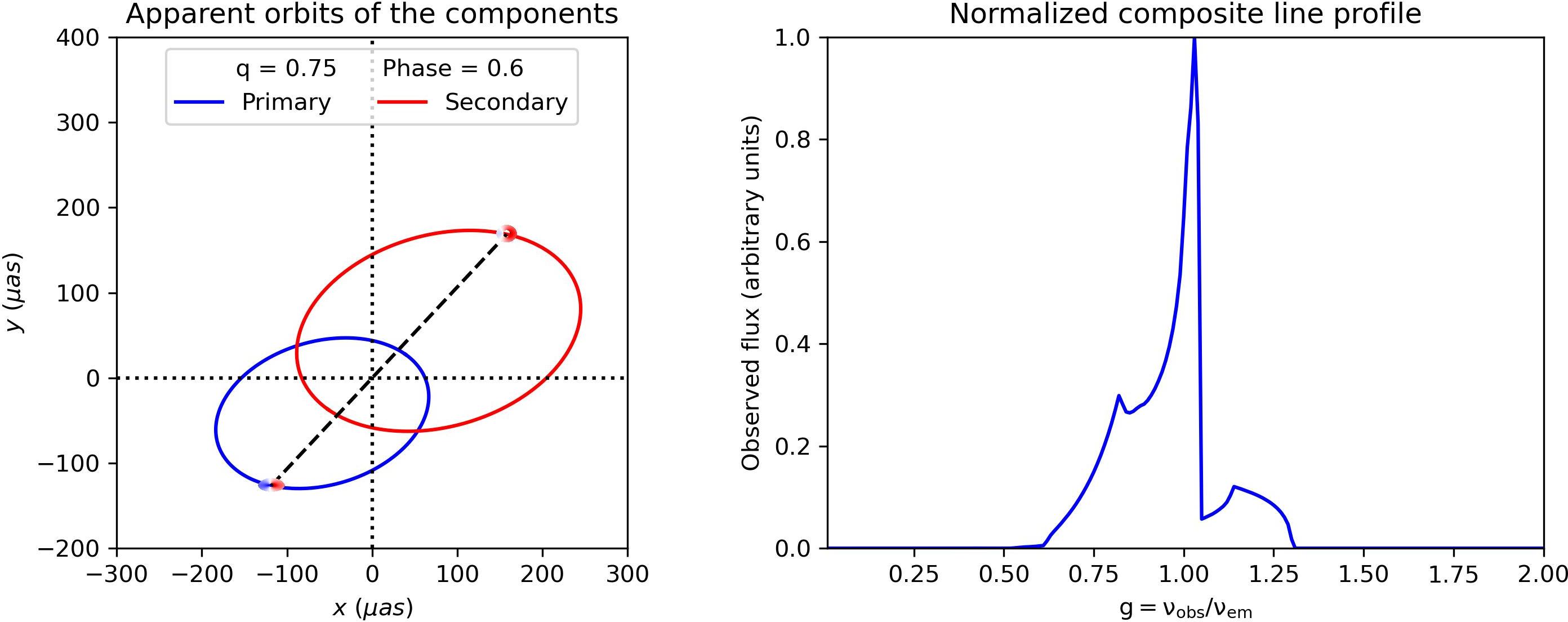}
\hfill
\includegraphics[width=0.49\textwidth]{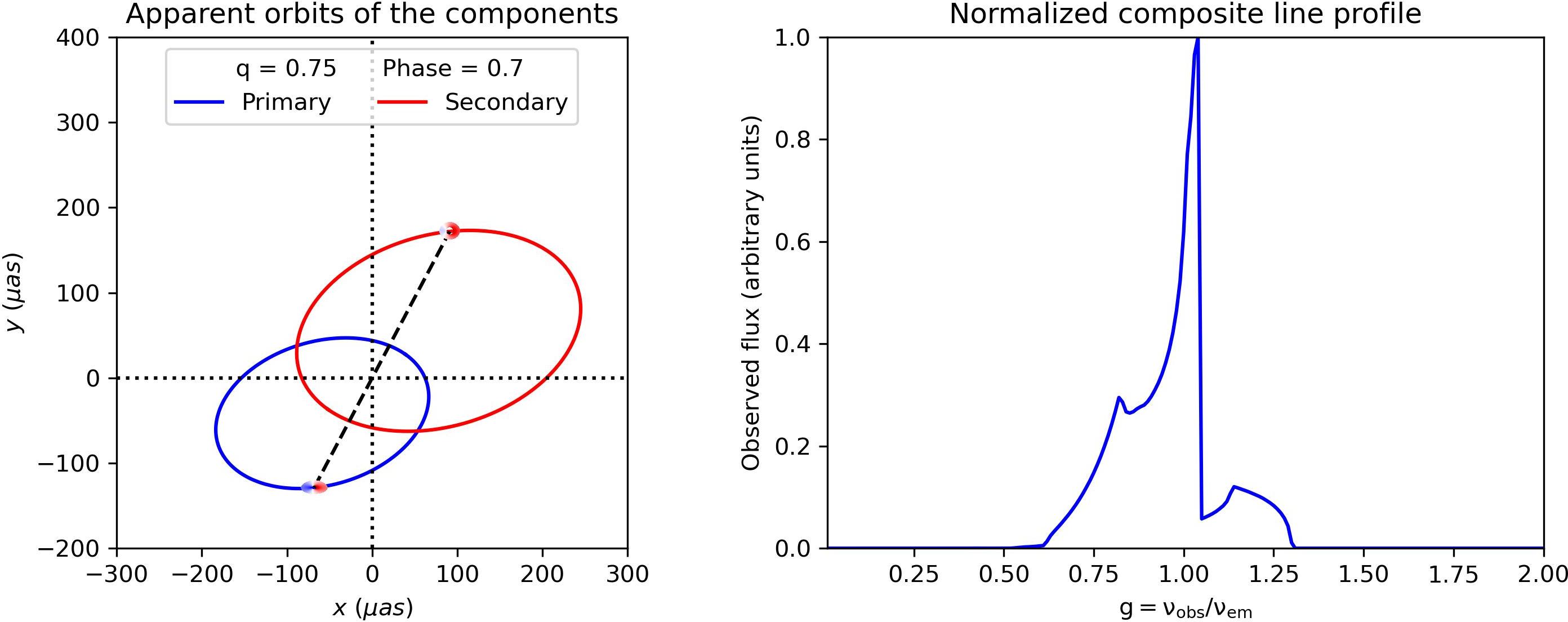}\\
\includegraphics[width=0.49\textwidth]{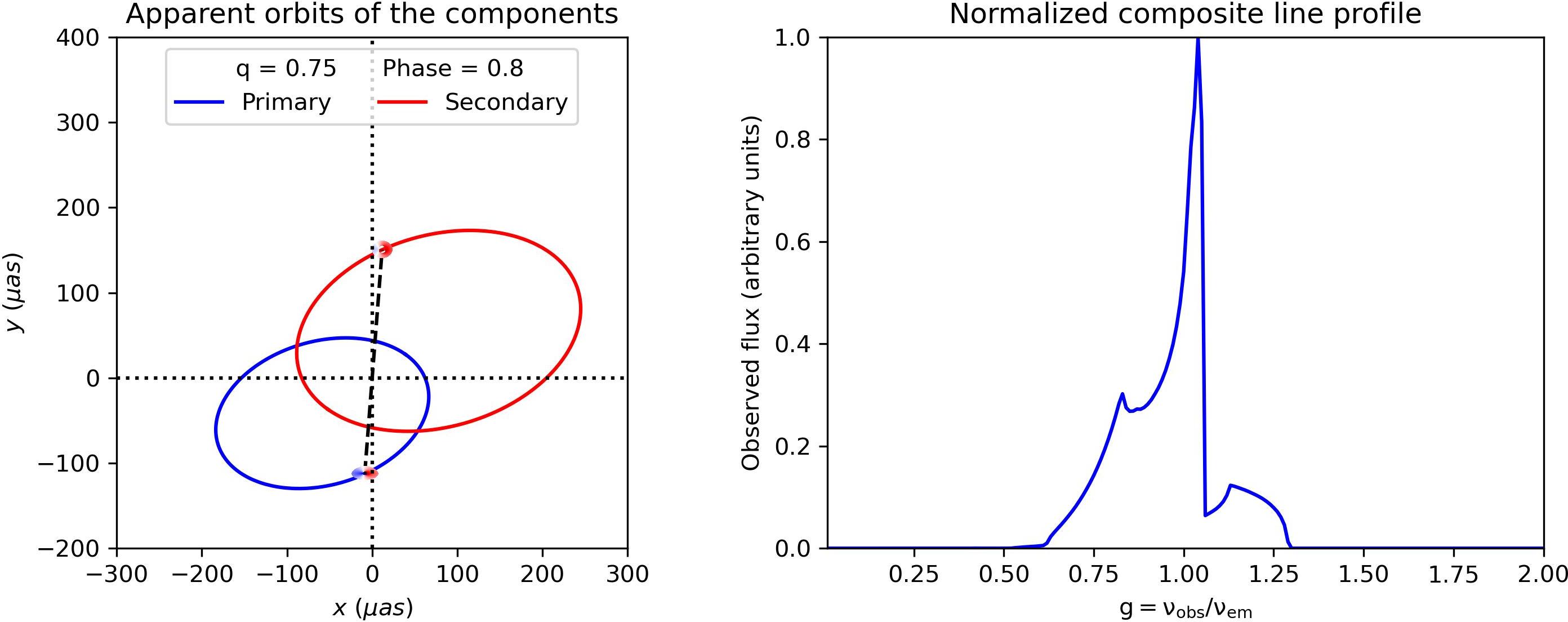}
\hfill
\includegraphics[width=0.49\textwidth]{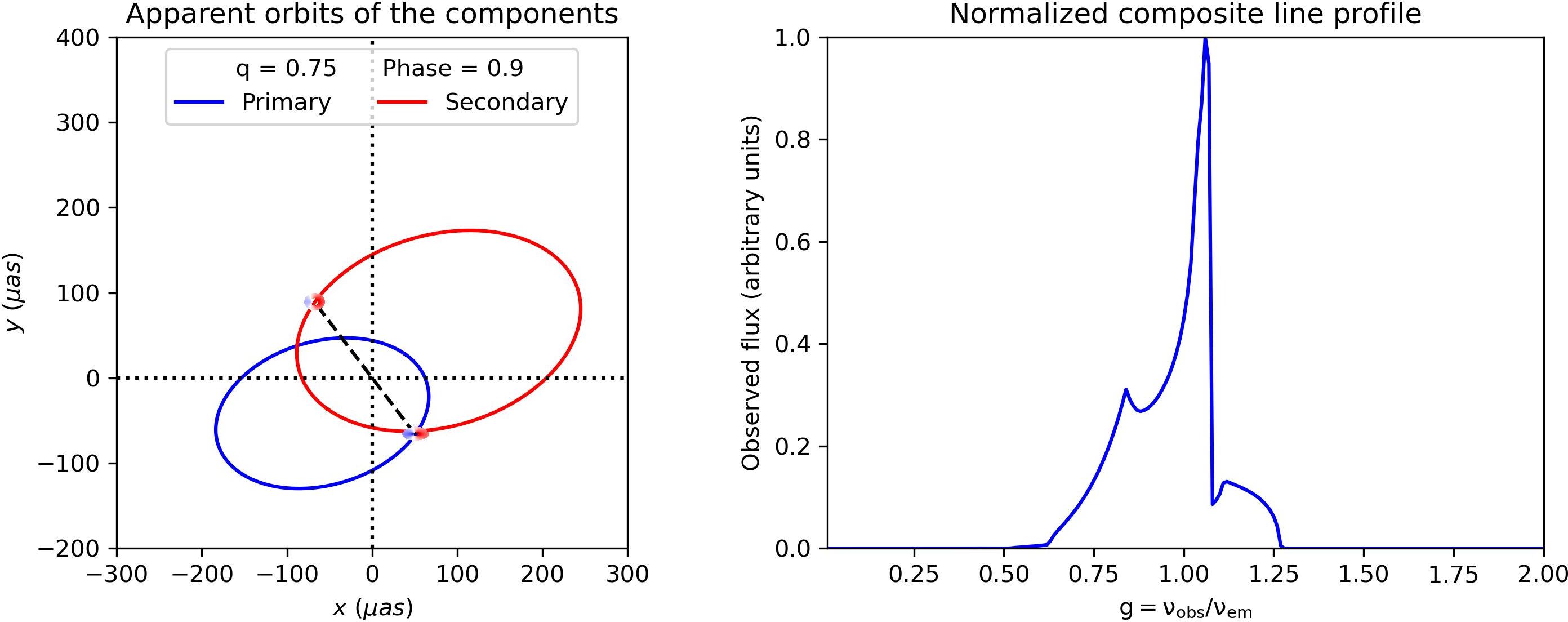}
\caption{10 simulated orbital phases of a SMBHB with disk 1 around the primary and disk 2 around the secondary
(left panels), as well as the corresponding composite Fe K$\alpha$ line profiles (right panels).}
\label{fig:disks12}
\end{figure*}

\begin{figure*}[ht!]
\centering
\includegraphics[width=0.49\textwidth]{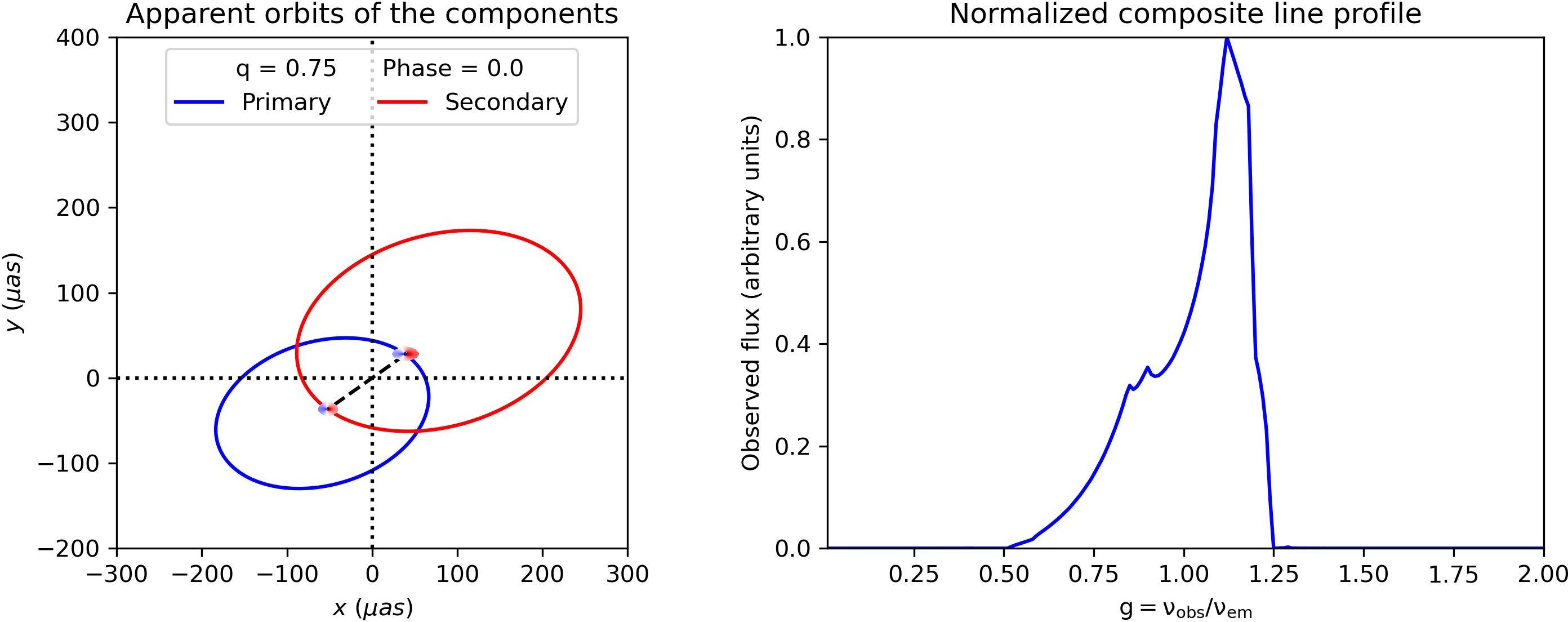}
\hfill
\includegraphics[width=0.49\textwidth]{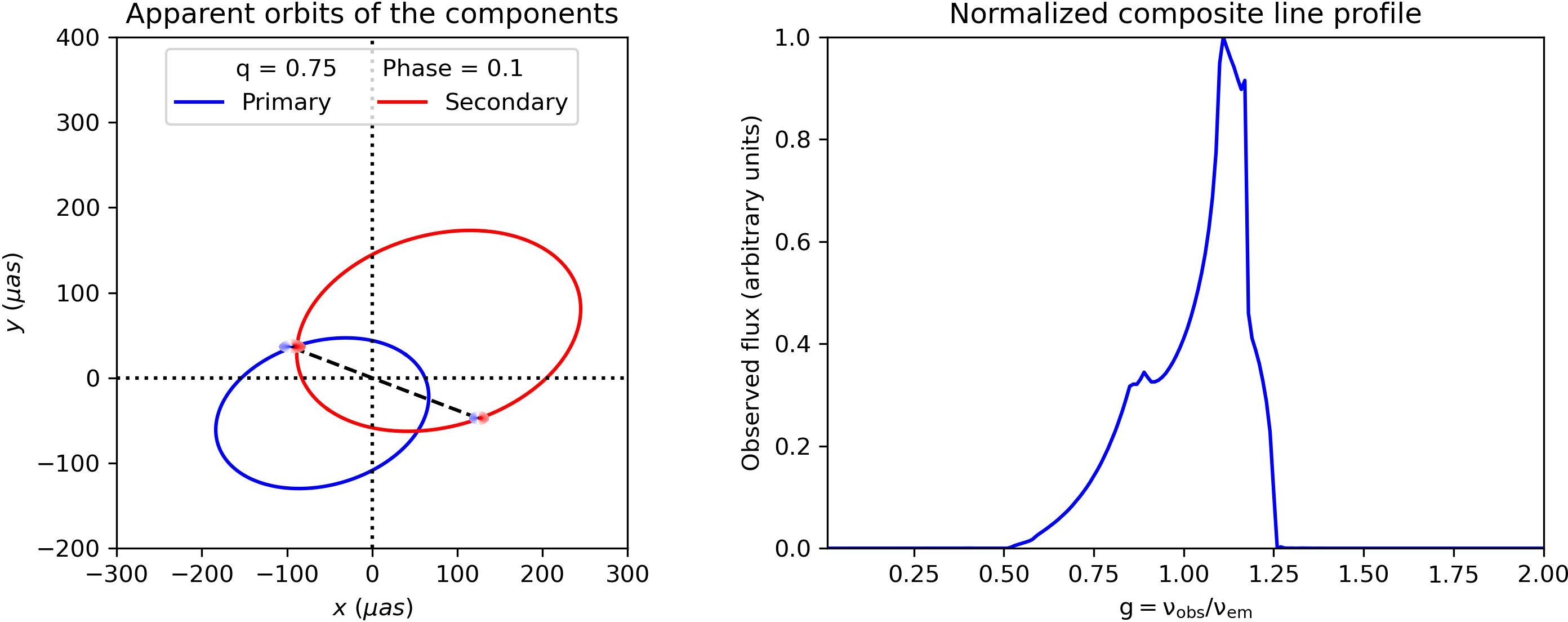}
\includegraphics[width=0.49\textwidth]{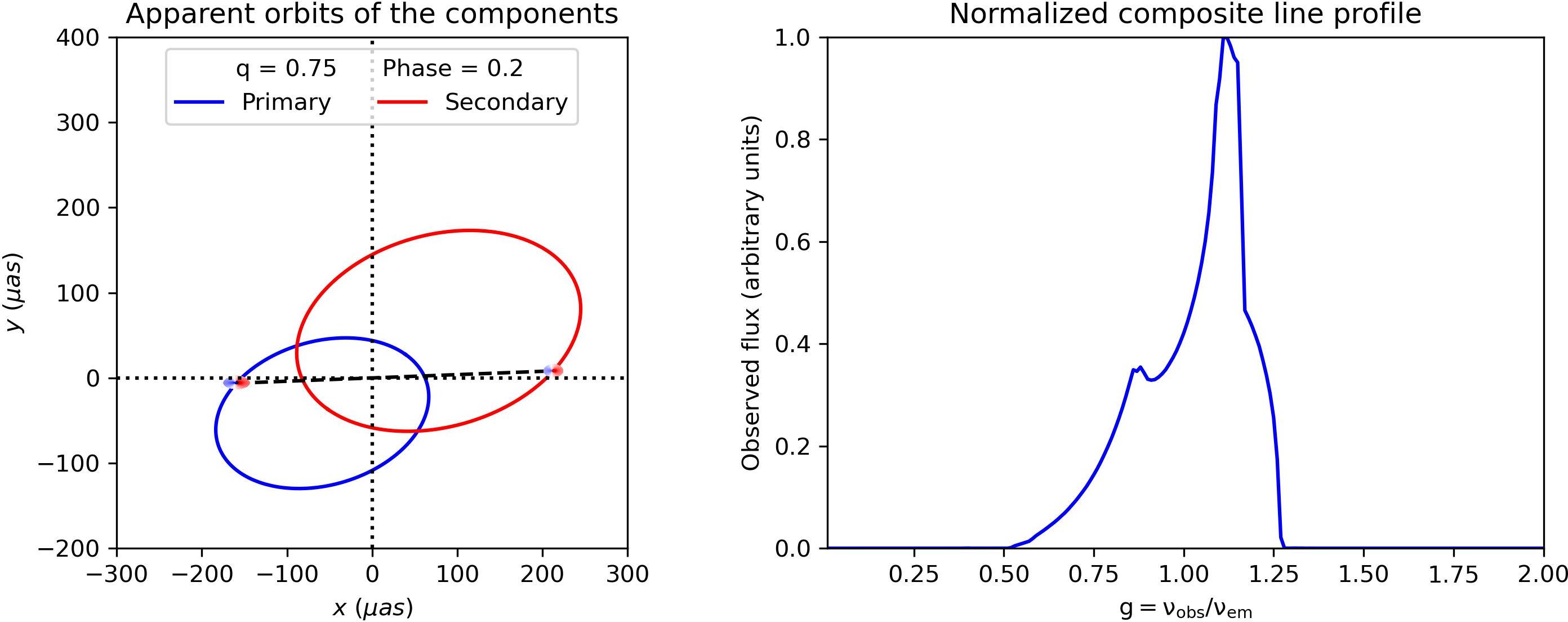}
\hfill
\includegraphics[width=0.49\textwidth]{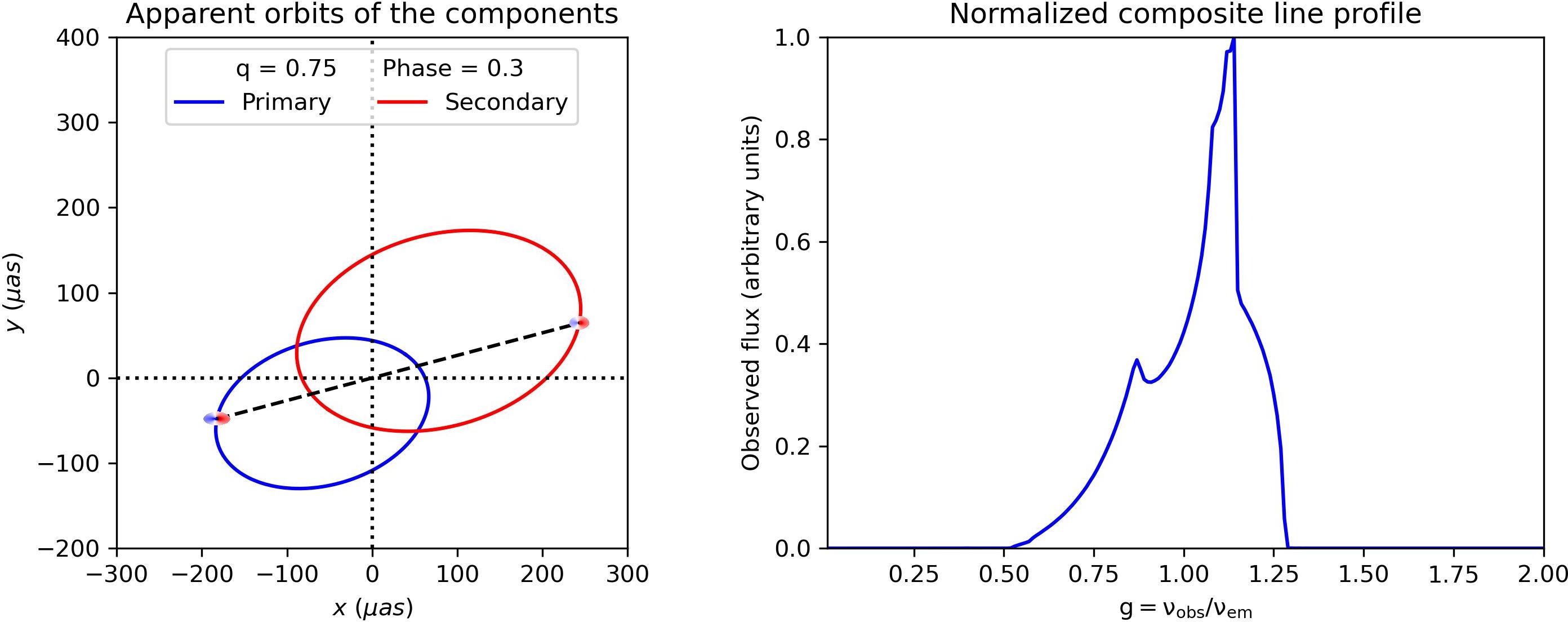}\\
\includegraphics[width=0.49\textwidth]{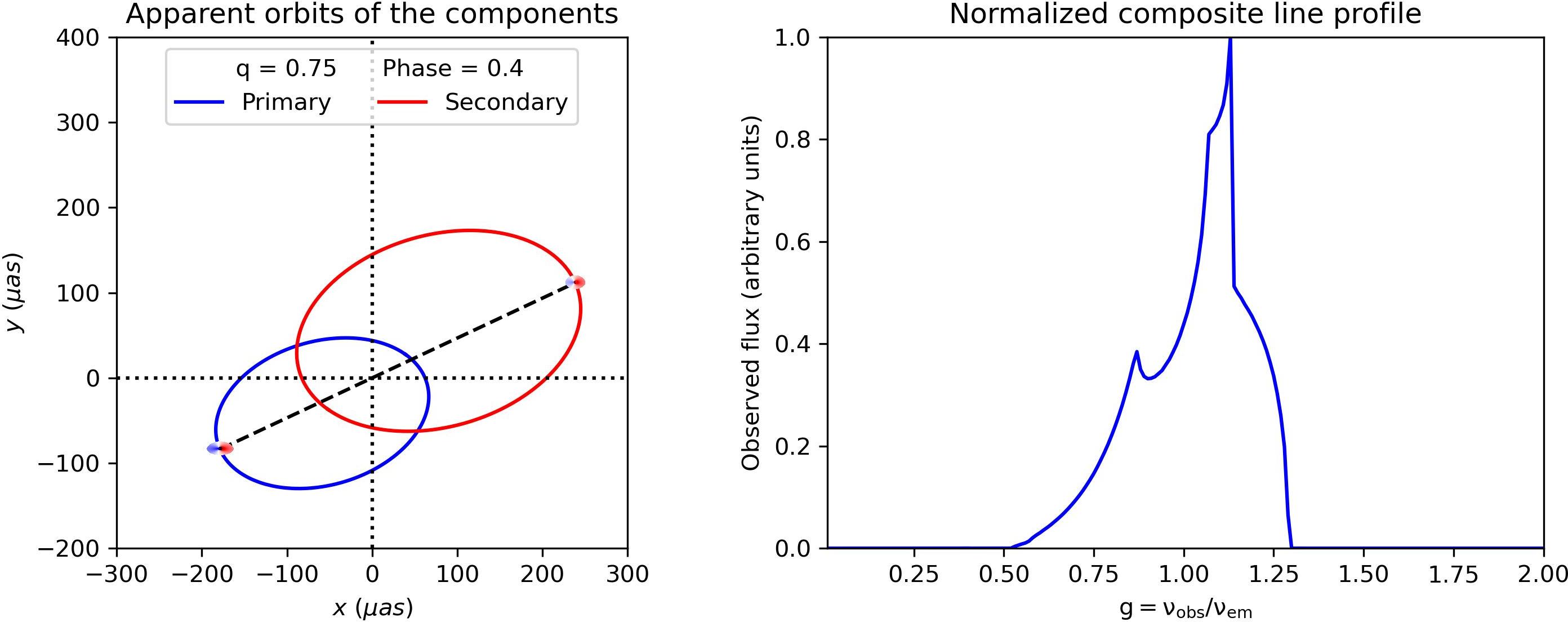}
\hfill
\includegraphics[width=0.49\textwidth]{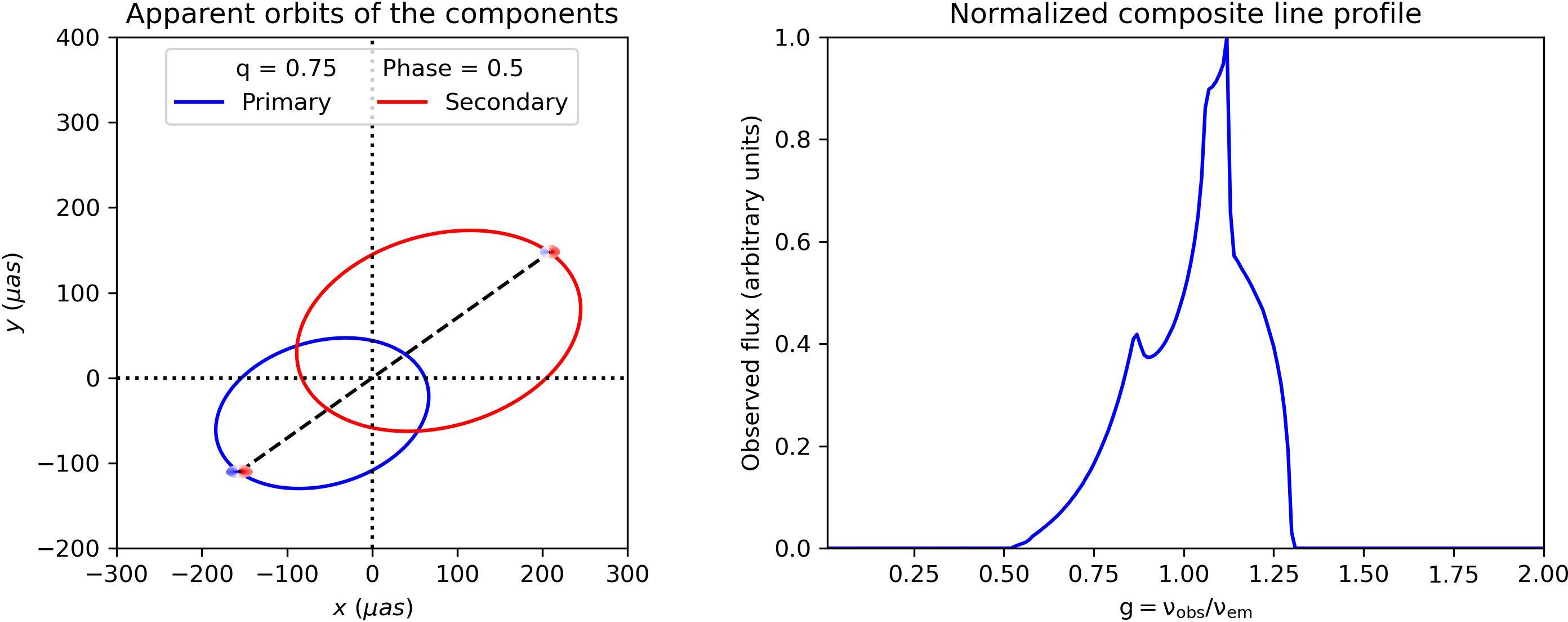}
\includegraphics[width=0.49\textwidth]{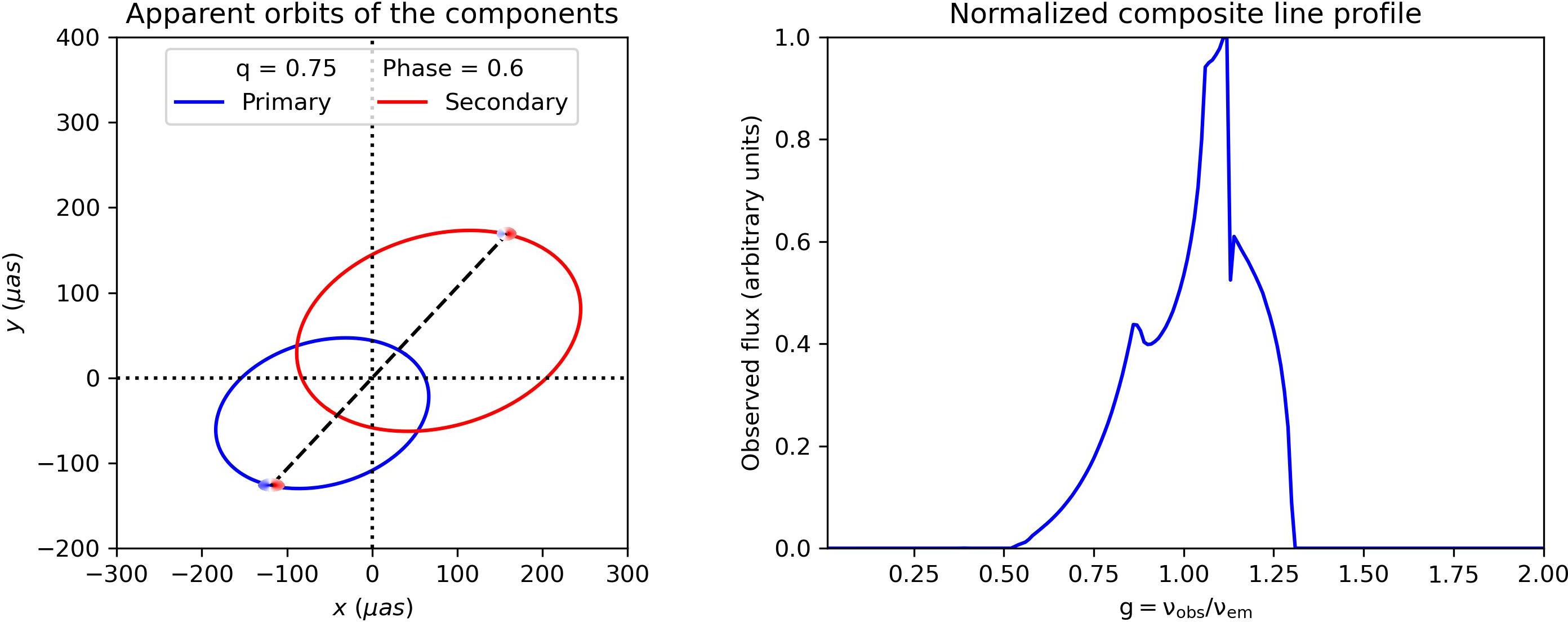}
\hfill
\includegraphics[width=0.49\textwidth]{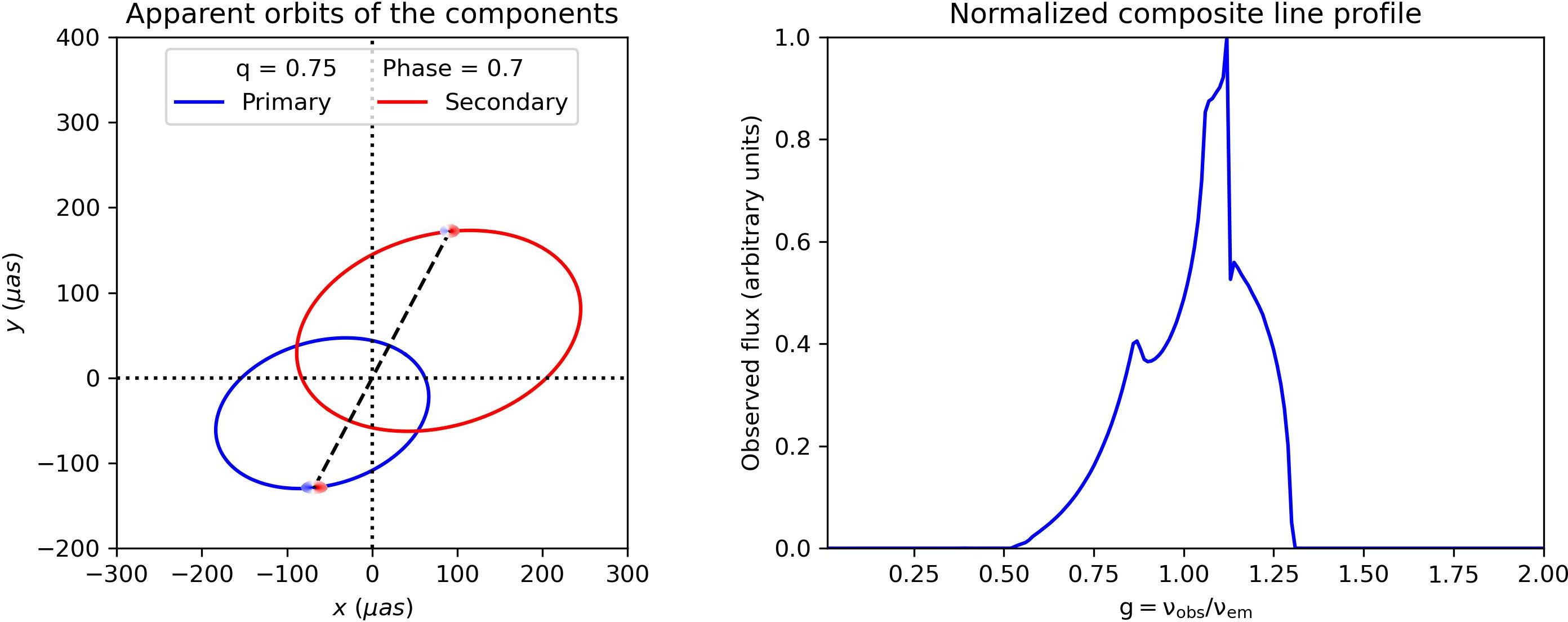}\\
\includegraphics[width=0.49\textwidth]{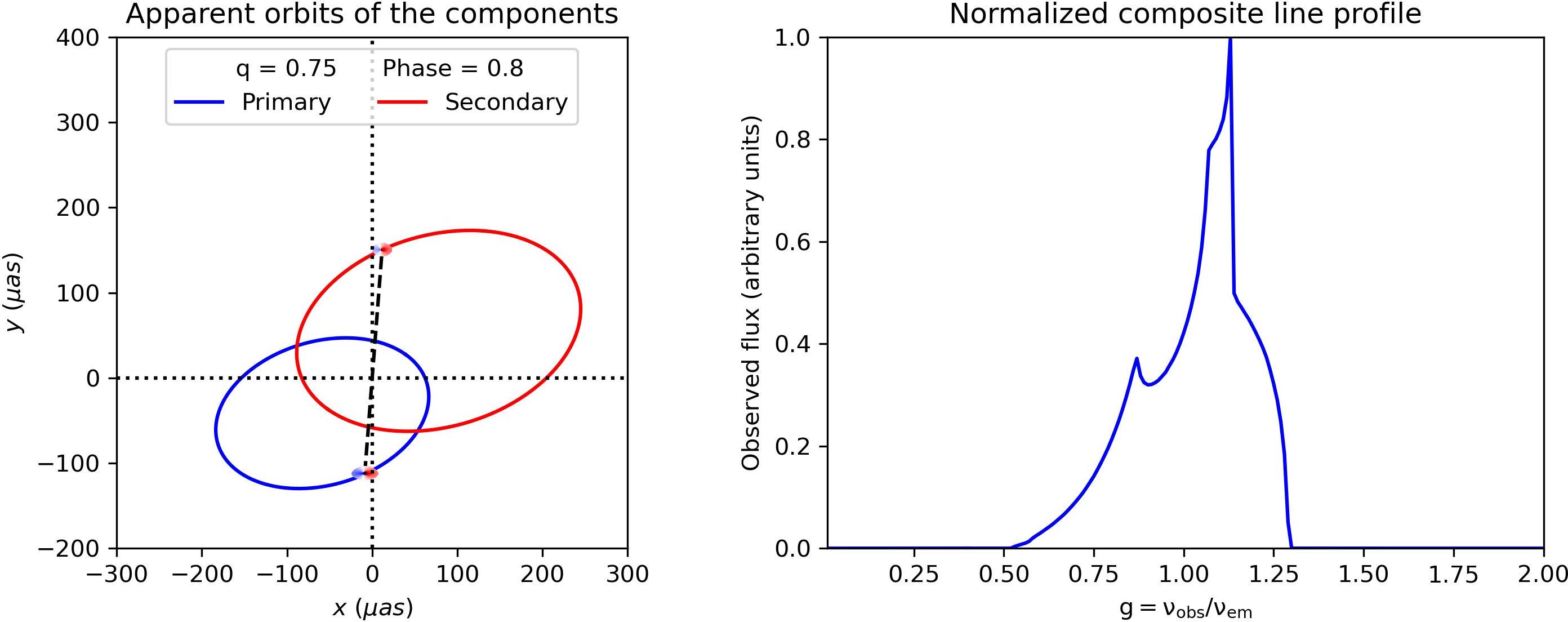}
\hfill
\includegraphics[width=0.49\textwidth]{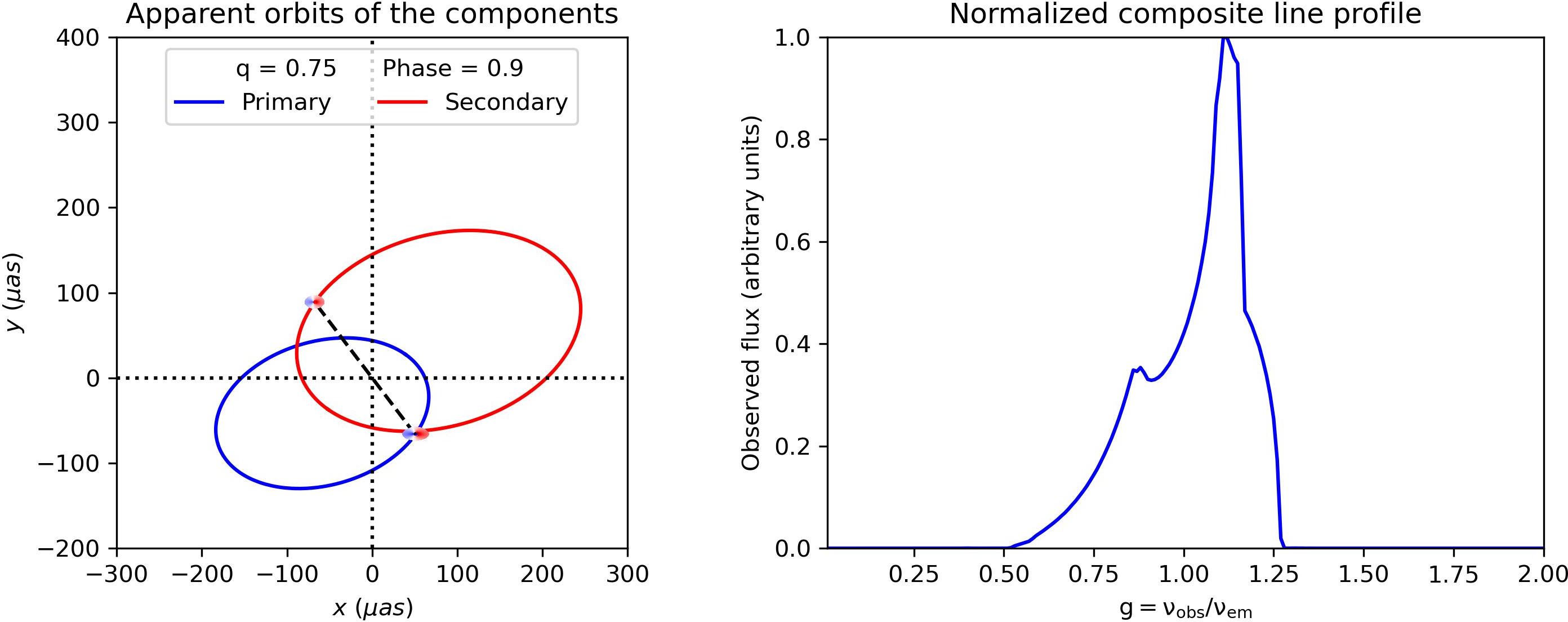}
\caption{The same as in Fig. \ref{fig:disks12}, but for disk 1 around the primary and disk 4 around the secondary.}
\label{fig:disks14}
\end{figure*}

\begin{figure*}[ht!]
\centering
\includegraphics[width=0.49\textwidth]{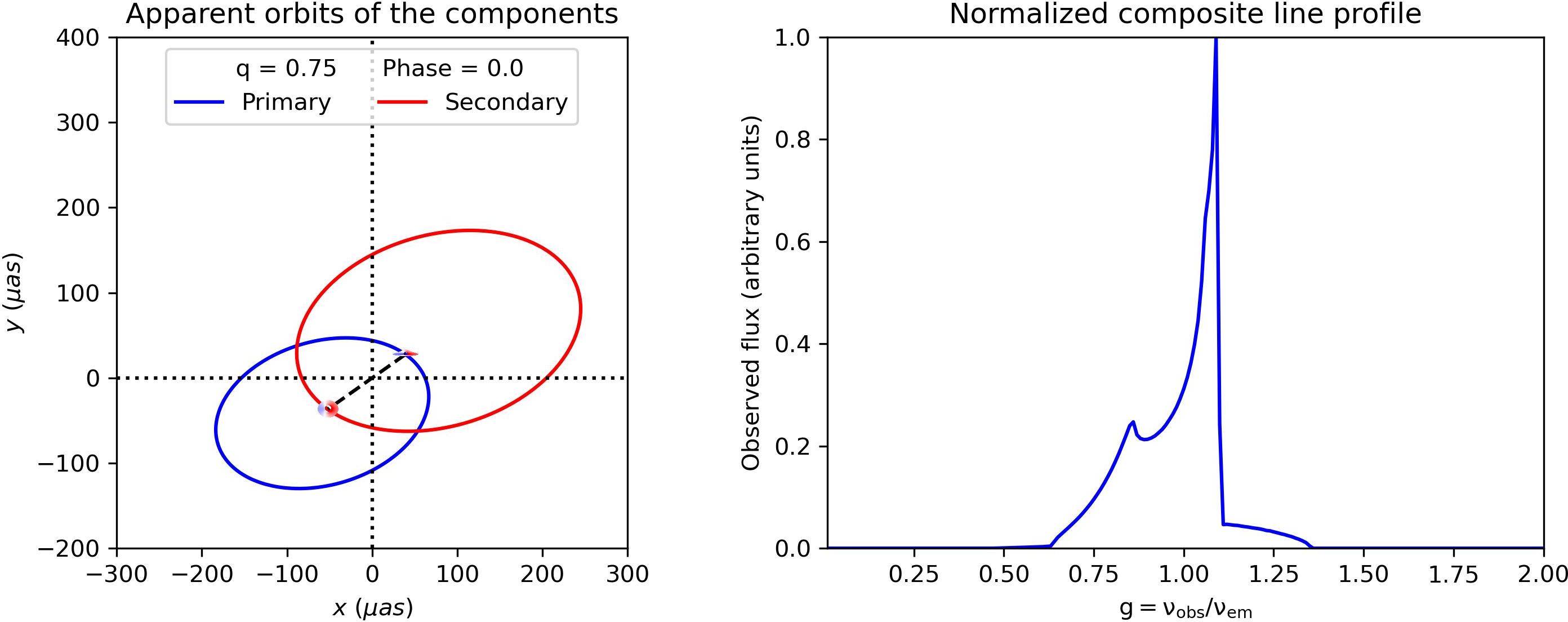}
\hfill
\includegraphics[width=0.49\textwidth]{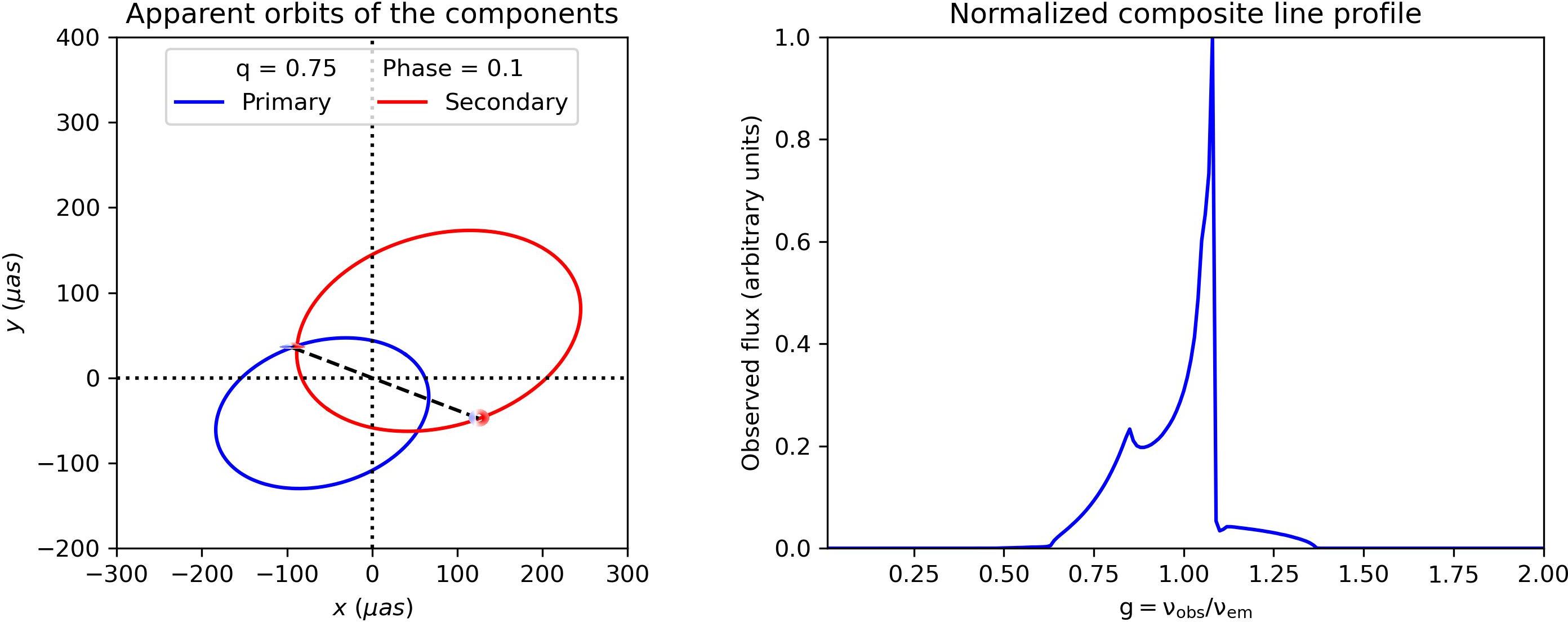}\\
\includegraphics[width=0.49\textwidth]{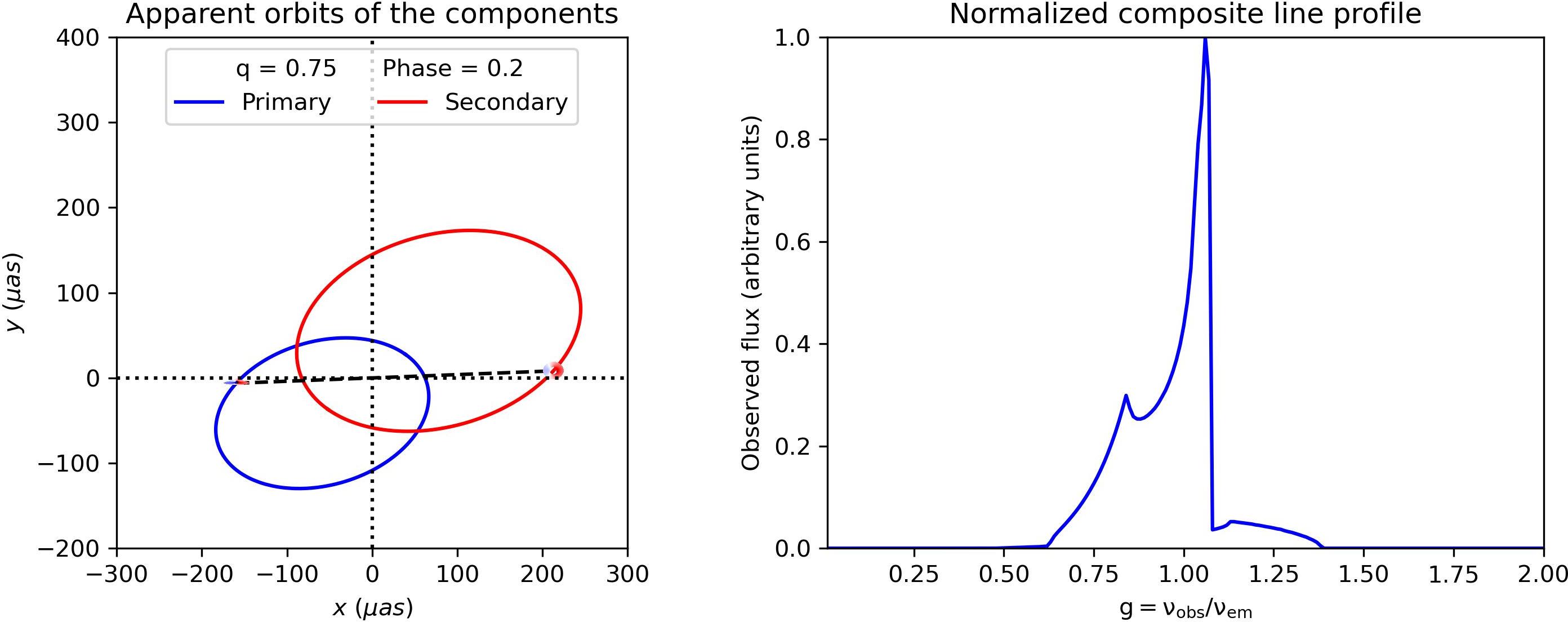}
\hfill
\includegraphics[width=0.49\textwidth]{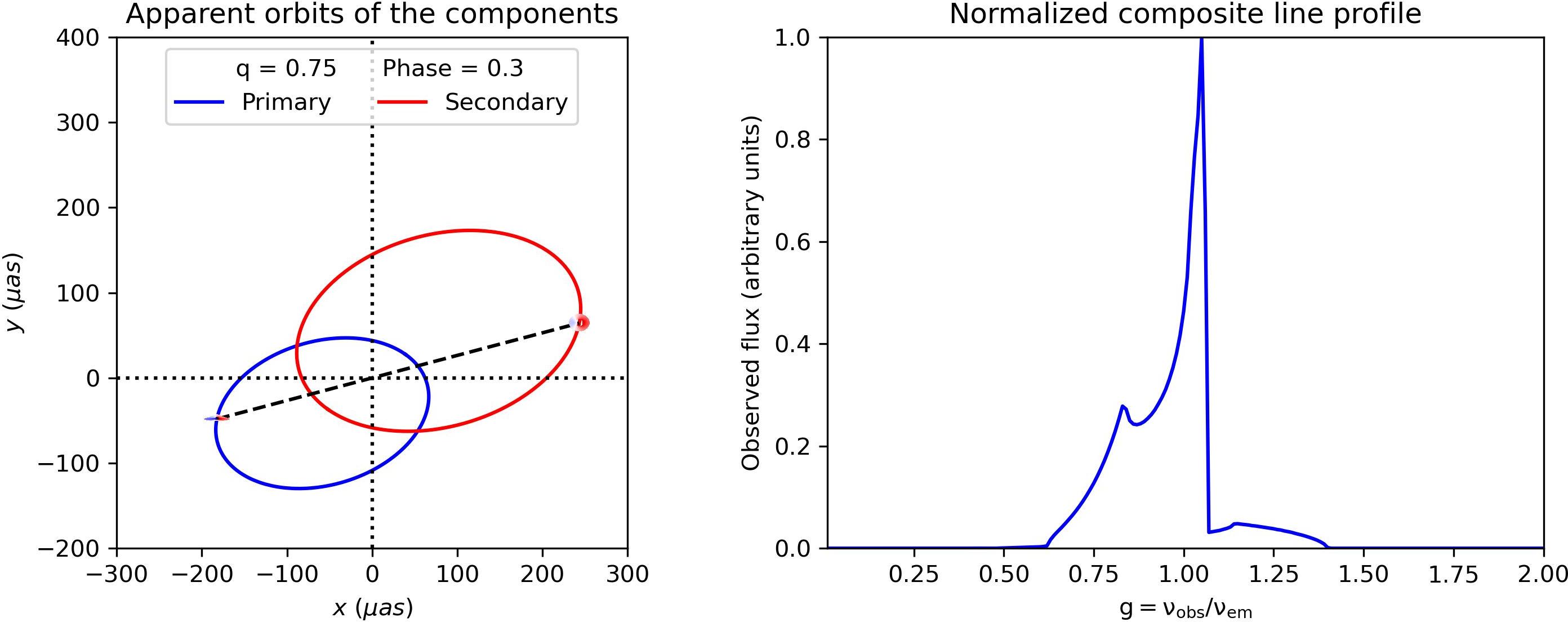}
\includegraphics[width=0.49\textwidth]{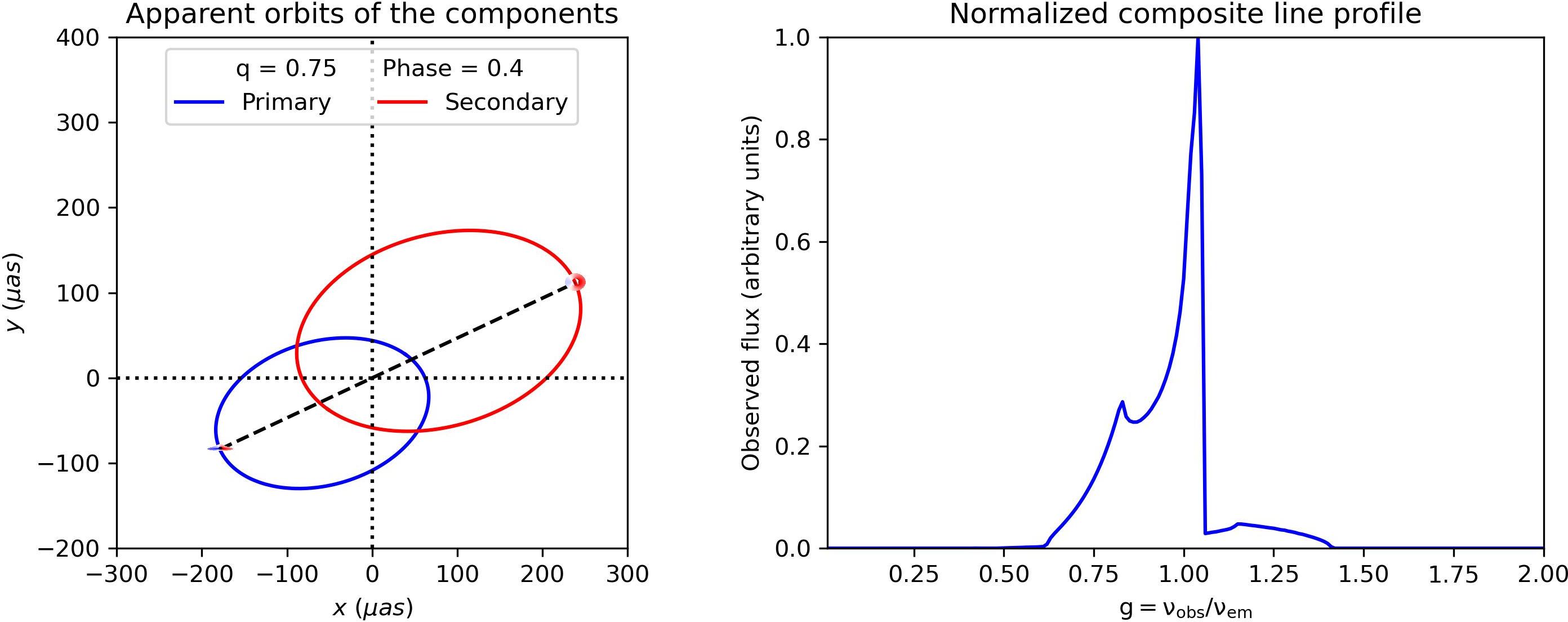}
\hfill
\includegraphics[width=0.49\textwidth]{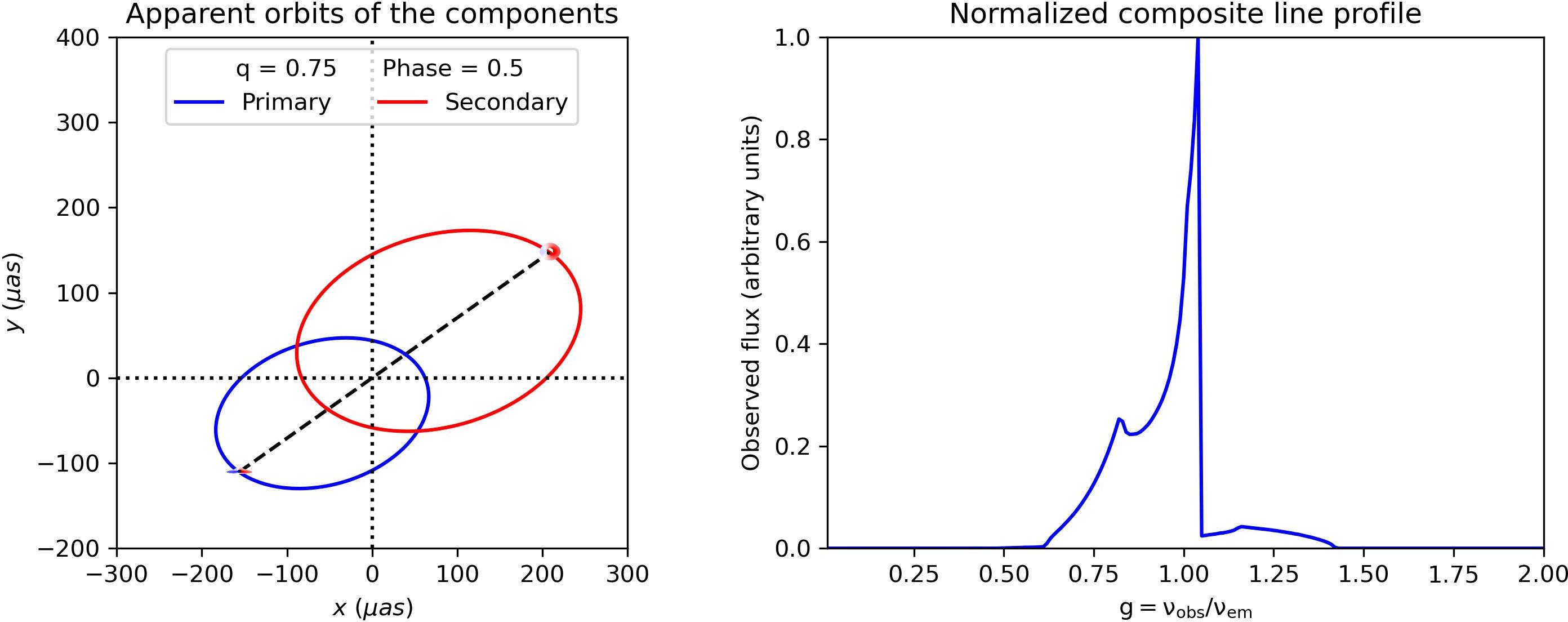}\\
\includegraphics[width=0.49\textwidth]{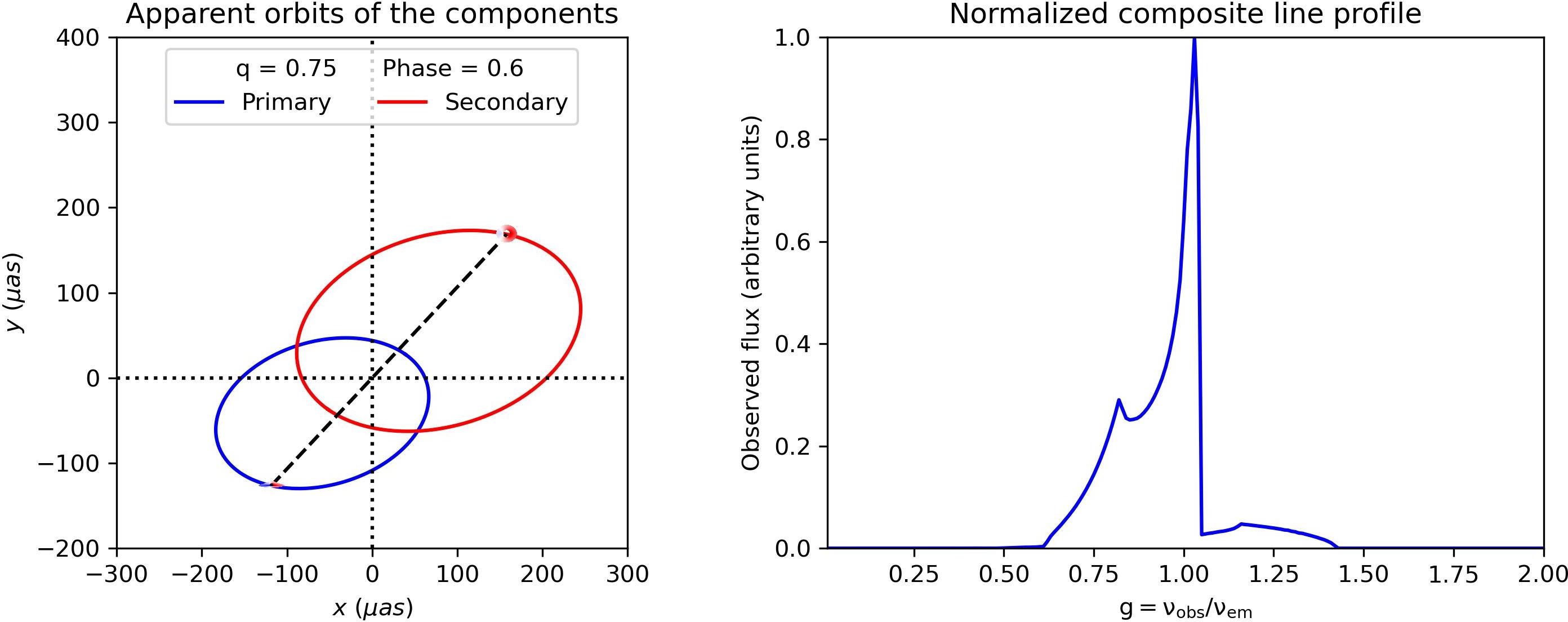}
\hfill
\includegraphics[width=0.49\textwidth]{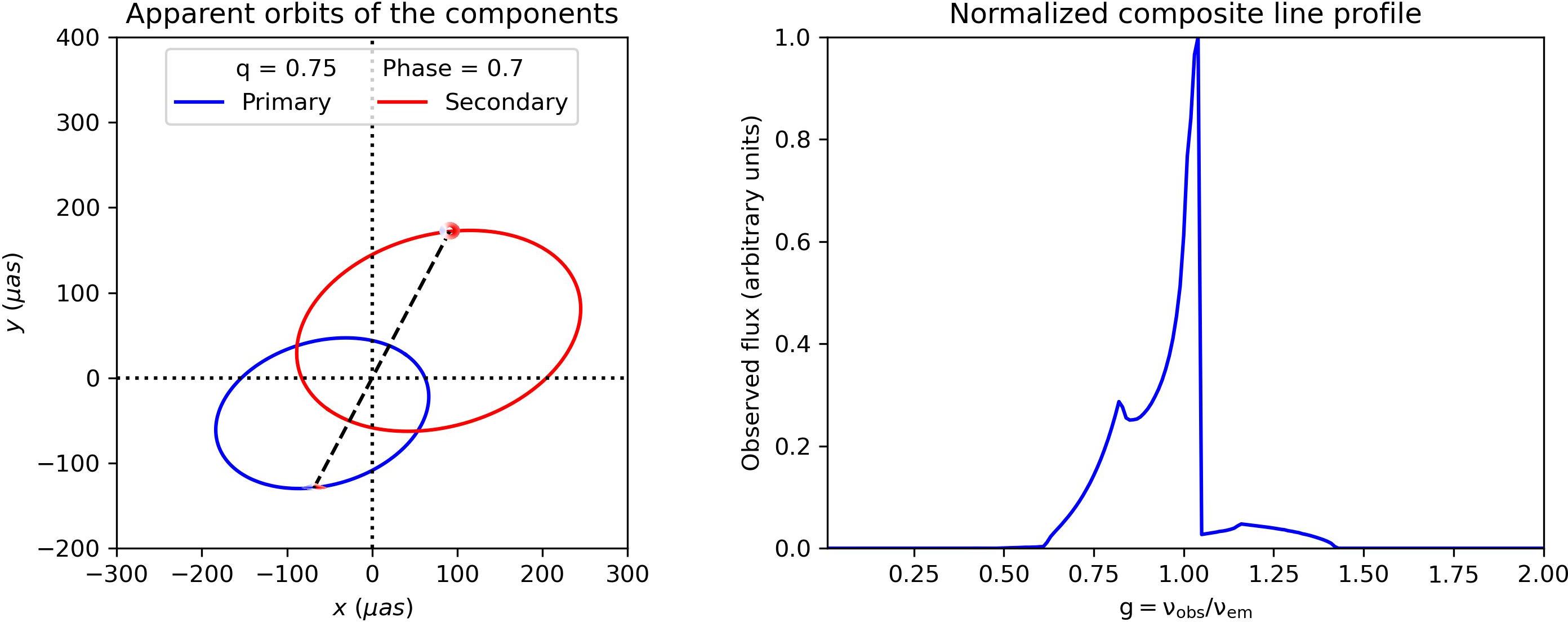}
\includegraphics[width=0.49\textwidth]{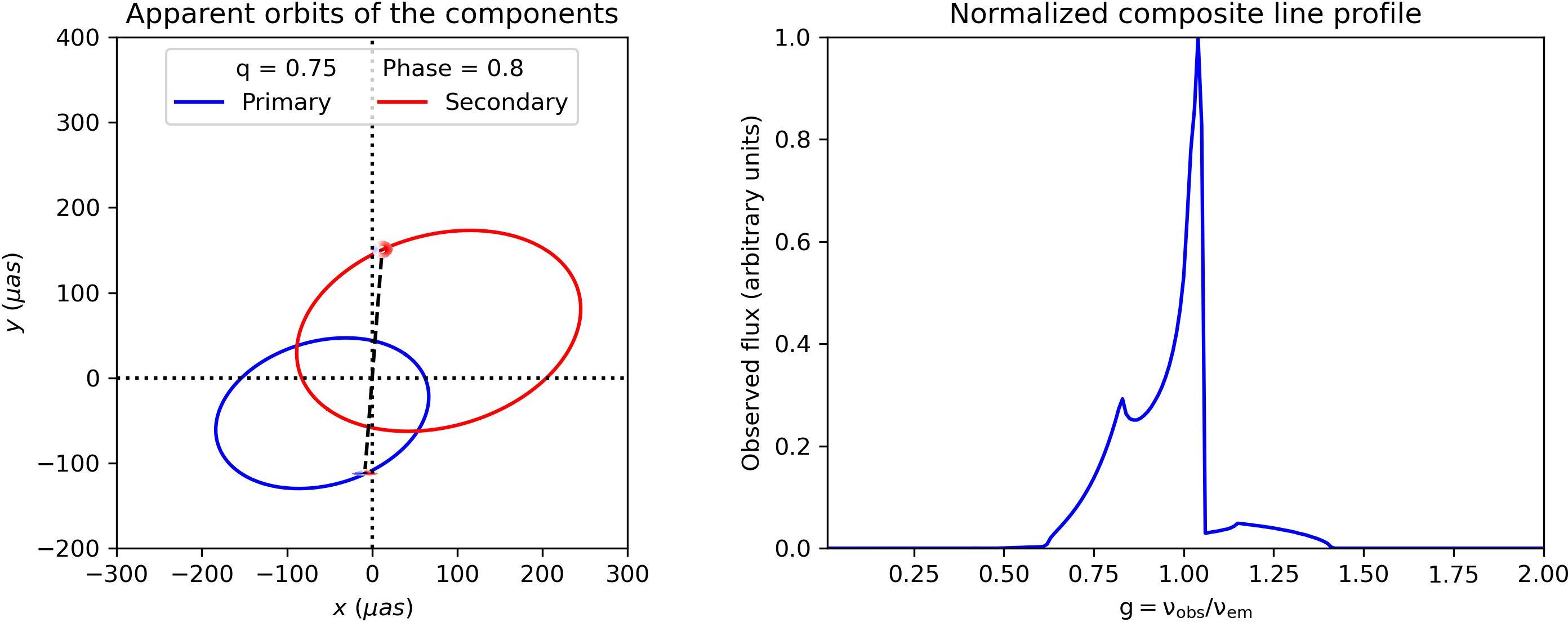}
\hfill
\includegraphics[width=0.49\textwidth]{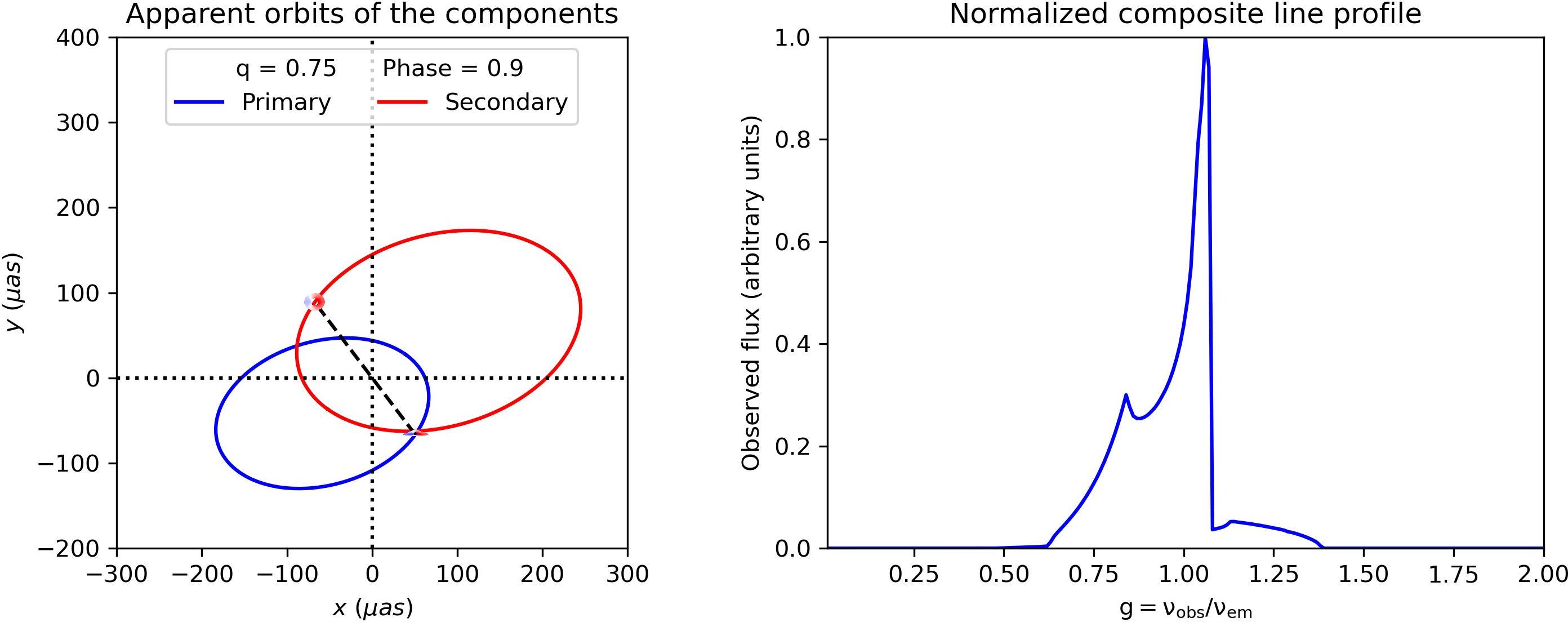}
\caption{The same as in Fig. \ref{fig:disks12}, but for disk 3 around the primary and disk 2 around the secondary.}
\label{fig:disks32}
\end{figure*}

\section{Results}

\subsection{Assumed SMBHB parameters and orbital elements}
\label{sec:smbhb}

We take into account the results of some previous studies \citep[e.g.][and references therein]{peter04,jova12,jova14,jova20} in order
to choose the parameters for modeling the accretion disks around the primary and secondary SMBHs, as well as their orbits.
We assumed a large mass of the primary SMBH and a small angular diameter distance to the binary system of SMBHs \citep[because the appearance of double relativistic Fe K$\alpha$ lines and periodic X-ray variability are expected to be detected from very massive and cosmologically nearby SMBHBs][]{sesa12}: $m_1=5\times 10^8\ M_\odot$ and $D_A = 16\;\mathrm{Mpc}$ \citep[corresponding to $z\approx 0.004$ according to the latest cosmological parameters by][]{pl18}: $H_0 = (67.4 \pm 0.5)\,\mathrm{km \, s^{-1}\, Mpc^{-1}},\, \Omega_m = 0.315 \pm 0.007,\, \Omega_\kappa=0.001 \pm 0.002$.
For the mass ratio of the SMBHB we adopted the value $q= m_2/m_1=0.75$, corresponding to the mass of the secondary of $m_2=3.75\times10^8\ M_\odot$, and we also assumed that the system's radial velocity of the center of mass is $\gamma=0\;\mathrm{km/s}$. For the Keplerian orbit of the studied SMBHB we adopted the following orbital elements: $a=5156.5\;\mathrm{AU}=0.025\;\mathrm{pc},\;e=0.6,\;i=45^\circ,\; \Omega=0^\circ,\;\omega=45^\circ$,
in which case the corresponding orbital period, according to Eq. (\ref{eqn:period}), is $P=12.52\;\mathrm{yr}$.

\subsection{Assumed parameters of the accretion disks}
\label{sec:disks}

We assumed the following inner and outer radii of the accretion disks emitting in the optical band:
$R^1_{in} = 0.1434\;\mathrm{mpc}$, $R^1_{out} = 1.2237\;\mathrm{mpc}$, $R^2_{in} = 0.1893\;\mathrm{mpc}$, $R^2_{out} = 1.4725\;\mathrm{mpc}$. Although in our simulations of optical emission we adopted the same inner radii as for X-ray band, these inner regions of disk radiate strongly only in X-ray band and do not contribute significantly to the optical emission, which mostly originates from much larger radii.

Regarding the assumed size of the X-ray emitting regions, as well as other disk parameters, we took into account the results of some theoretical and observational studies, according to which the X-ray radiation is emitted from the innermost regions of the accretion disks and its spectrum depends on the distance to the central SMBH \citep[see e.g.][]{shak73,reyn97,ball05,nand07,patr12}.
Therefore, in order to study the X-ray variability of SMBHBs, we used the following four models (denoted as disk 1-4) of the accretion disks emitting in the X-ray band, as described by the SMBH spin $J$, observed viewing angle (disk inclination) $\theta_{obs}$, inner ($R_{in}$) and outer ($R_{out}$) radius of the X-ray emitting region, and spectral index $p$ of power law emissivity:
\begin{itemize}
\item disk 1: $J = 0.1,\; \theta_{obs} = 60^\circ,\; R_{in} = 6.0 R_g,\; R_{out} = 50 R_g,\; p =  -2.5$ ;
\item disk 2: $J = 0.1,\; \theta_{obs} = 30^\circ,\; R_{in} = R_{ms} = 5.67 R_g,\; R_{out} = 20 R_g,\; p =  -2.5$ ;
\item disk 3: $J = 0.1,\; \theta_{obs} = 80^\circ,\; R_{in} = 6.0 R_g,\; R_{out} = 50 R_g,\; p =  -2.5$ ;
\item disk 4: $J = 0.1,\; \theta_{obs} = 45^\circ,\; R_{in} = 6.0 R_g,\; R_{out} = 50 R_g,\; p =  -2.5$ .
\end{itemize}

It should be noted that, although SMBH spin is commonly denoted by $a$, here we dentoted it with $J$ (which is a commonly
used label for angular momentum) in order to avoid confusion with the semimajor axis of the orbit. However, one should
bear in mind that here $J$ represents angular momentum normalized to SMBH mass, so that $0 \le J \le 1$ (i.e. $J$ is spin).

All four disks models of SMBHs are assumed to be slowly rotating, with the same small spin: $J = 0.1$. We choose 4 different disk inclinations $\theta_{obs} = 60^\circ$, $\theta_{obs} = 30^\circ$,
$\theta_{obs} = 80^\circ$ and $\theta_{obs} = 45^\circ$ in case of disk 1, disk 2, disk 3 and disk 4, respectively. Inner radii of disks 1, 3
and 4 of SMBHs are fixed to $R_{in} = 6.0\, R_g$ (where  $R_g=GM/c^2$ is the gravitational radius of the SMBH with mass $M$)
and their outer radii are also assumed to be the same: $R_{out} = 50\, R_g$. In case of disk 2 inner radius is
$R_{in} = R_{ms} = 5.67\, R_g$ (where $R_{ms}$ is the radius of the marginally stable orbit) and outer radius is:
$R_{out} = 20\, R_g$. In all 4 cases disk emissivity indices are: $p = -2.5$.

We simulated the X-ray radiation from all four disk models, and in Fig. \ref{fig:disks} we presented the obtained simulated disk
images (colored according to the energy shifts $g$ and the observed fluxes $F_{obs}$), as well as the corresponding simulated non-normalized Fe K$\alpha$ line profiles for all 4 disk models. As it can be seen from Fig. \ref{fig:disks}, disk inclination has the strongest influence on the width of the resulting line profiles. However, one should note that spin is fixed to $J = 0.1$ but it can also significantly affect the line widths and profiles, especially their right wings due to the gravitational redshift which is stronger for larger spins \citep[see e.g.][]{jova12}. It should be possible to detect these effects of both inclination and spin on line width using current and future generation of X-ray observatories, such as Imaging X-ray Polarimetry Explorer (IXPE), Advanced Telescope for High Energy Astrophysics (Athena) and X-Ray Imaging and
Spectroscopy Mission (XRISM). This is also in a good agreement with our previous findings \citep{jova12,jova14,jova20} and represents the main reason why we assumed and presented these particular 4 disk models.

\subsection{The variability of line profiles in optical band}

We generated spectrum for specific parameter set (Table \ref{tab:param_set}) and present the results in Figures \ref{fig:broad_optical} and \ref{fig:HB_line_continuum}.

\begin{table}[ht!]
\centering
\begin{tabular}{c|c|c|c|c|c}
$m_1\;\mathrm{[M_\odot]}$ & $m_2\ \mathrm{[M_\odot]}$ & $R\mathrm[pc]$ & $i\mathrm{[^o]}$ & $e$ & $\omega\mathrm{[^o]}$\\[0.5ex]
\hline
$5\times10^8$ & $3.75\times10^8$ & 0.025 & 45 & $0.6$ & 30
\end{tabular}
\caption{Parameters values for line profiles in optical band. The description of enlisted parameters are given in previous text.}
\label{tab:param_set}
\end{table}

In this spectra we included just strong lines, which can clearly reflects influence of orbital phases on their shift and shape. We take $Ly\alpha[\lambda1215]$, $\mathrm{CIV}[\lambda1549]$, $\mathrm{CIII}[\lambda1908]$, $\mathrm{MgII}[\lambda2798]$,
$\mathrm{H}\alpha[\lambda6564]$, $\mathrm{H}\beta[\lambda4861]$, $\mathrm{H}\gamma[\lambda4341]$, $\mathrm{H}\delta[\lambda4102]$.

As we can see, the presented spectrum is changing during the orbit, and that is mainly due to the variations in continuum flux and in line shape. As we proposed in the basic description of our model, continuum is generated by accretion disks; therefore, orbital dynamics strongly influence the distribution of continuum emission. Within the proposed parameter set, radial velocities of components are opposed, resulting in simultaneous continuum shifts of particular components toward blue and red spectrum parts. Therefore, cumulative continuum flux, which is the sum of continuum emission from both components, depends on their mass ratio $q$ and orbital phase.

On the other hand, simulated spectrum lines vary in intensity for a number of reasons. First, the line intensity directly depends on the ionization conditions in the BLR, particularly the BLR sizes, which are determined by radiation heating from accretion disks. Therefore, it is expected that line shape, especially line intensity, periodically changes in accordance with the orbital dynamics of the system.
Second, due to the superposition of radiation from BLRs of both BHs with different radial velocities and a stationary circumbinary BLR, cumulative line profiles are perturbed in intensity and width. 
This is particularly significant in cases of moderate and high eccentric orbits. Additionally, in those cases, mutual interaction \citep[see]{popo21} between components can be significant, resulting in observable variation in the continuum emission and line shape.
As we mentioned above, the Keplerian motion of a broad line emitting gas around each black hole is taken as the dispersion of the corresponding Gaussian assumed as line profiles. In the case of cBLR, we assume that width is influenced by the Keplarian motion of emitting clouds around the total mass of the binary. However, the cBLR does not follow the dynamical motion of the binary system and has no radial component of motion connected with the dynamical effects.

\subsection{The variability of line profiles in X-ray band}

In order to simulate the electromagnetic signatures in the Fe K$\alpha$ lines emitted from the nearby SMBHBs and to study
variability in the line profiles in the X-ray band, we performed simulations of the X-ray radiation from
the SMBHB with parameters and orbital elements presented in \S\ref{sec:smbhb}, during one orbital period.
The outer radii of the X-ray emitting regions in the disks around the primary are adopted to be: $R_{out}=50\;R_g$ which,
for the assumed mass of the primary of $m_1=5\times 10^8\ M_\odot$, corresponds to $\approx 250\;\mathrm{AU}$, and in the case
of a SMBHB located at distance of $16\;\mathrm{Mpc}$, it corresponds to an angular size of $\approx 15\;\mu\mathrm{as}$.

Fig. \ref{fig:vrad} presents the obtained radial velocities (left) and redshift factors (right) of the components in the studied SMBHB. It can be seen from
the left pannel of this figure that the radial velocity of the secondary SMBH can go almost to 10,000 km/s, and it means that it can induce significant Doppler shift in the X-ray radiation from its accretion disk (see the right panel of the same figure).

We simulated the X-ray radiation from the studied SMBHB during different orbital phases along its orbit using the calculated radial velocities
shown in the left panel of Fig. \ref{fig:vrad}. For that purpose we assumed the following three different combinations of disk models around the SMBHB components:
\begin{enumerate}
\item
disk 1 around the primary and disk 2 around the secondary;
\item
disk 1 around the primary and disk 4 around the secondary and
\item
disk 3 around the primary and disk 2 around the secondary.
\end{enumerate}

The obtained simulated Fe K$\alpha$ line profiles emitted from the disks around the primary and secondary component, as well as their normalized composite profile, are presented in Fig. \ref{fig:periapo} for each of the three combinations and during the orbital phases 0.0 and 0.5, i.e while the SMBHB components are in pericenter (left panels) and apocenter (right panels), respectively. As it can be seen from Fig. \ref{fig:periapo}, the redshifts induced by the radial velocities can significantly affect the resulting composite line profiles. This influence depends on orbital phase of the binary system, as well as on the disk models around its components, and it is the most noticeable in the second case, when disk 1 is around the primary and disk 4 around the secondary.

In order to study whether the orbital motion of the studied SMBHB has a similar effect on its optical and X-ray emission, we compared the simulated normalized profiles of the composite H$\beta$ and Fe K$\alpha$ lines, emitted at the same orbital phases. These comparisons are presented in Fig. \ref{fig:hbvsfa} for all three disk combinations around the SMBHB components and during their pericenter (left panels) and apocenter orbital phases (right panels). As it can be seen from Fig. \ref{fig:hbvsfa}, the redshifts induced by the radial velocities have noticeable effects on both H$\beta$ and Fe K$\alpha$ lines. It seems that in the latter case, this influence is much more prominent, and thus it is easier to detect the imprints of the orbital motion in the X-ray radiation. However, one cannot expect to find similar features in both the optical and X-ray line profiles. We should note here that we assumed the Gaussian shapes for BLR emissions of each component. We should note here that we assumed the Gaussian shapes for BLR emissions of each component so that SMBH mass strongly affects the H$\beta$ FWHM, while the dynamical effects have significant influence on the line intensity ratios and the shape of the composite H$\beta$ line. On the other hand the Fe K$\alpha$ line of each component is calculated taking into account relativistic effects and composite line. It would be better to take the H$\beta$ disk-like emission for each BLR, but here we investigate the influence of dynamical effects in the case when the widths of broad lines reflect the BH masses and therefore, in the first approximation, one can conclude about the dynamical effects on the H$\beta$ line profiles. In future investigations, it would be better to take disk-like BLR emission for each BH component and ring-like emission of cBLR.

In Figs. \ref{fig:disks12}-\ref{fig:disks32} we presented 10 simulated orbital phases (denoted by 0.0, 0.1,\ldots ,0.9) of the studied SMBHB for each of three disk combinations for the primary and the secondary (left panels), as well as the corresponding composite Fe K$\alpha$ line profiles (right panels). In all studied cases, the disk images are colored according to the total redshift factor $g_{tot}$ and are presented in the left parts of each panel in Figs. \ref{fig:disks12}-\ref{fig:disks32}, while their right parts show the corresponding simulated composite Fe K$\alpha$ line profiles.

By inspection of Figs. \ref{fig:disks12}-\ref{fig:disks32}, we can conclude that our model of SMBHB induces either the appearance of the ripples
in the cores of the simulated composite Fe K$\alpha$ line profiles, or variability in their wings.
The profiles of the simulated lines strongly depend on the assumed disk models around both primary and
secondary SMBHs, as well as on the particular orbital phase. By comparison between the three cases presented in Figs. \ref{fig:disks12}-\ref{fig:disks32}, it can be
seen that the strongest variability appears in the red wings of the composite line profiles, while the ripple effect in the
line cores are noticeable only in the second case (Fig. \ref{fig:disks14}). Since the line profiles for different orbital phases
differ from each other in all three cases, it can be deduced that the main cause of the line variability is the orbital motion of the SMBHB components.
In our previous results \cite{jova14}, we demonstrated that the mass ratio $q$ also has a strong influence on the strength of
such variability.

The presented results of our simulations demonstrate that SMBHBs can induce observable variability in their Fe K$\alpha$ line profiles.
This gives the possibility of detecting such effects in the observed spectra of SMBHB candidates, especially if future X-ray detectors improve their spectral resolution and signal-to-noise ratio (S/N).

\section{Discussion and Conclusions}

In this paper, we investigate the influence of the SMBHB dynamics on the broad line profiles in the X-ray (i.e., the Fe K$\alpha$ line) and UV/optical (taking H$\beta$ broad line) in order to find some similar behavior on the line profiles in two considered wavlength bands. For the simulation of the UV/optical spectra, we used a non-hydrodynamical model developed as described in \cite{popo21}. For simulation of the Fe K$\alpha$ line profile, we used the model presented in \cite{jova14}. The simulations were performed using the same parameters for the SMBHB systems, and we compared the line shapes in the optical and X-ray spectra.

Presented method could be used for researching variability of spectral lines in SMBHB, by studying double-peak line profiles or line shifts induced by radial velocity components. We showed that model can generate those profiles, which give us possibility to compute used parameter sets. However, there are some uncertainties and limitations invoked by free parameters in the model. First of all contributions to $\mathrm{H_\beta}$ line from cBLR is artificially reduced to be between 5 to 10\% of the calculated emission. This action is justified by the fact that this region is at larger distance and consequently less heated then surrounding BLRs of the components. 
Additionally, chosen parameters set include high eccentric orbits ($e=0.6$ in all computations), which results in significant radial velocity and consequently visible double peaked lines in computed spectra. On the other hand, when using more circular orbits, double peaked line profile is rare, since it is masked by line width, producing more Gaussian shaped profile. However, in such case shift of the combined line profile can indicate possible SMBHB candidate. 
Also, we neglect the aperiodic effects which could be generated by accretion disk winds and jets. In some specific cases, when generated on both components with time scales higher then $P_{orb}$, they could possible induce additional variability in optical or X-ray domain. Nevertheless, we have shown using this simple model, that spectroscopic detection is applicable for a number of SMBHB with high eccentricity and inclination angle.

Similar investigation with more details is given in \citep{nguyen16}, which showed that system parameters like component distance, mass ratio, eccentricity, and the degree of alignment of BLR disks can influence profile of generated lines. Also, \cite{Panda19} showed using CLOUDY code to model the BLR emission, that system orientation has significant influence on quasar optical spectra. 
In our paper we confirmed their conclusions, but also found that parameter list can be broadened with additional variables, which include ratio of emission contribution from BLRs and cBLR. Additionally, we find that modeling BLR dimensions with Roche lobes directly influence the velocity distribution in those regions in a way that with more compact systems, dimensions of the BLRs are decreased and consequently increase line widths. Therefore in that cases we expect that produced lines rather show Gaussian then double peaked profiles, with possible line shift.

Our results showed that SMBHBs could cause a specific, but different variability of the optical spectral lines and Fe K$\alpha$ line, leaving potentially detectable imprints in their profiles. It seems that a more prominent change in the line profiles can be seen in the case of the UV/optical broad lines than in the Fe K$\alpha$, especially if we have in mind a low spectral resolution in the X-ray detectors. However, the change in the Fe K$\alpha$ profile seems to be very promising since we expect a very fast change in the line profile, which is not the case in the UV/optical part.

Additionally, in this study, we found that the masses of the BH components and their mutual distance have the most significant influence on the produced spectra in both wavelength bands since they define orbital dynamics as well as the efficiency of the radiation processes involved. A special case is considered when discussing low/high mass ratio systems, which cause rearrangement of BLRs in the SMBHB and subsequent consequences, as discussed in the paper \citep{popo21}.
However, we found that eccentricity $e$, inclination $\theta$ and orbital orientation angle $\omega$ can also have significant influence on the produced lines. For example, if the orbit is more eccentric it will cause higher radial velocities in a part of the orbit when components are in the perihelion position. Therefore, one can expect that if the orientation angle is $\omega=0$ radial velocities will be highest, and we expected double peaked line profile to be produced. However, as our simulations shows, line intensity in that case is significantly decreased, since in the compact binary systems mutual interaction is for this configuration also highest, causing significant perturbation in BLRs and consequently reduction in line intensity. This effect can be more significant if combined with some arbitrary angles $\omega$ and/or $\theta$.

We should mention here that double-peaked line profiles can be observed in the case of the BLR emission, which is coming from a single black hole. In the case of the inclined disk-like BLR, an observer can detect receding and approaching parts of the emitting gas, where receding emission will contribute to the red and approaching to the blue part of a line profile that will reflect in the double peaked line profile shapes \citep[see][]{chen89,pop04,dosSantos23,ochmann24}. However, sometimes a double-peaked line profile can indicate the presence of a supermassive binary system, but also single-peaked profiles can be formed in a binary system \citep[see e.g.,][]{pop12,pop20}.

Concerning our simulations of the same SMBHB dynamical parameters and different emissions in X-ray and UV/optical spectral bands, we can outline the following conclusions:
\begin{itemize}
\item SMBHB imprints depend on the orbital phase of the system that reflects in both (X and optical) spectral regions. The line profiles of Fe K$\alpha$ and H$\beta$ could be used for reconstructing the Keplerian orbits of the components in the observed SMBHBs in both spectral bands. Moreover, such signatures in the optical and X-ray line profiles of the observed SMBHBs could be used as a tool for the detection of these objects as well as for studying their properties. The broad line profiles emitted from SMBHB systems show bumps and shoulders of two peaks, but it depends on the phase, and in some cases, the profiles do not show any indication that there are SMBHB.
\item
There are not similar line profiles in the Fe K$\alpha$ and H$\beta$ broad lines, even though the configuration of the SMBHB is taken to be the same. This is causing the difference in the region of the broad line origin since the optical spectrum is emitted in the complex BLR (formed from moving and non-moving emission gas) that is much further away from accretion disks, which emit in the Fe K$\alpha$ line. 

\end{itemize}

As final conclusions, the answer to the question of whether there is similar variability in the broad Fe K$\alpha$ and H$\beta$ spectral line shapes caused by SMBHB that can be easily recognized is negative. We cannot expect similar line-shape variability in these two lines. Even H$\beta$ may show the shape that is usual for an AGN, while Fe K$\alpha$ may show an unusual profile.

\section*{Acknowledgments}
The Authors acknowledge the support of Ministry of Science, Technological Development and Innovations of the Republic of Serbia through the Project contracts No. 451-03-66/2024-03/200002, 451-03-65/2024-03/200122 and 451-03-66/2024-03/200017.

\bibliographystyle{jasr-model5-names}
\biboptions{authoryear}
\bibliography{ms}

\end{document}